\documentclass[a4paper,fleqn]{cas-sc}

\usepackage[numbers]{natbib}
\usepackage{subcaption}
\usepackage{amsmath}
\usepackage{url}
\usepackage{lineno}
\usepackage{nccmath}

\def\tsc#1{\csdef{#1}{\textsc{\lowercase{#1}}\xspace}}
\tsc{WGM}
\tsc{QE}
\tsc{EP}
\tsc{PMS}
\tsc{BEC}
\tsc{DE}
\usepackage{enumerate}
\begin{document}
\let\WriteBookmarks\relax
\def\floatpagepagefraction{1}
\def\textpagefraction{.001}
\shorttitle{S-shaped neutron guide simulations}
\shortauthors{APS Souza, LP de Oliveira and FA Genezini}

\title [mode = title]{Monte Carlo simulations of the S-shaped neutron guides with asymmetric concave and convex surface coatings}

\author[1]{APS Souza}
\cormark[1]
\ead{alexandre.souza@ipen.br}

\address[1]{Instituto de Pesquisas Energéticas e Nucleares, IPEN/CNEN, Av. Prof. Lineu Prestes, 2242 – Cidade Universitária – CEP 05508-000 São Paulo – SP – Brazil}

\author[1]{LP de Oliveira}


\author[1]{FA Genezini}


\cortext[cor1]{Corresponding author}
\begin{abstract}
During the last decades, neutron beam transportation has been a well-known and established subject of designing proper neutron guides. However, sometimes unusual adaptation or adjustments are required out of original projects and after the operation beginning of facilities. An inter-center transfer of instrument locations also requires a new approach that is not necessarily described in the literature. Inside these situations, the use of S-shaped guides has not been fully discussed in the literature. From a geometrical analysis, we have developed a construction formalism of a minimal S-shaped guide by only considering the exclusion of the Line-of-Sight. The guide model has been studied through the wavelength cutoff and the neutron transport efficiency analysis. Here, Monte Carlo simulations using MCSTAS software has been applied. By intending to optimize these guide systems, simulations of this study also consider scenarios that have different supermirrors. A formalism to determine wavelength cutoff for unique and variable index guide systems has also been developed. There is a good correspondence between the theoretical cutoff wavelength values and the corresponding values obtained through Monte Carlo simulations. In addition, we have found specific configurations that combine efficient neutron transport and lower m-values on the convex surfaces of curved guides that form the S-shaped guide.
\end{abstract}


\maketitle

\section{Introduction}
 
The transportation of neutrons by different types of guides and components has always been a subject of intense scientific activity \cite{MaierLeibnitz1963,Mildner2008,romain2016}, since the first nuclear fission reactor in the 1940s \cite{1982} to the recent projects \cite{Perrotta2014,Souza2020}, and spallation sources \cite{Garoby2018}. The theories describing the transportation of neutrons were divided into two groups, the deterministic methods that lead to approximations to solve the Boltzmann equation \cite{lamarsh1966introduction} and the stochastic approach, based on the Monte Carlo method \cite{Lewis1993}.

The MCNP (Monte Carlo N Particle) code was a milestone in this scenario, enabling a sophisticated description of reactor cores with different geometries, compositions and particle transport, including neutrons, photons and the neutron-photon channel \cite{Goorley2014}. In the early 1990s, Nielsen and Lefmann created an open access software (MCSTAS), with a friendly interface capable of describing the transport of neutrons through guides and other optical components, making it possible a stochastic simulation of neutron scattering instruments by different materials \cite{Lefmann1999}.

An S-shaped neutron guide has been an element of studies recently \cite{Heinemann2015,DeOliveira2021,deOliveira2020}. The S geometry was the solution found for installing a SANS instrument at the \textit{Forschungs-Neutronenquelle} Heinz Maier-Leibnitz Laboratory from Munich (FRM-II), which has received instruments from the decommissioning of FRJ-2 reactor. In this case, there were different beam hole heights from the level of the instrument's sample site in both reactors \cite{Gilles2006}. In addition to the relative height, the S-guide provided a cutoff in the neutron spectrum, what is impossible to occur in a curved guide only. This geometry, therefore, allows efficient cutting of fast (epithermal) neutrons to a cold (thermal) source, something undesirable for neutron guides and Helium 3 detectors. More recently, this type of guide has been studied and relations among the radius of curvature ($\rho$), the length ($L$), and the super mirrors ($m$) have also been found \cite{deOliveira2020}. The index $m$ that characterizes the super mirrors corresponds to the ratio between the critical angle of the guide material and the critical angle of neutron reflection on a Nickel-58 surface.

In this context, the use of S-shaped guides is based on allocating instruments to different and pre-existent neutron beam holes and tubes. After the decommissioning of the research reactors, we could see many examples of instrument transfer to different facilities, e.g., the installation of the reflectometer SPATZ from HZB BER-II in the OPAL \cite{lebrun2016}. Otherwise, guide systems are designed and optimized just to avoid epithermal neutrons and gamma rays and guarantee a fine neutron flux at the instrument entrance, by means of curved guides or neutron filters for instance. In these terms, an S-shaped guide, which consist of two curved guides connected in different senses, has to exclude the system's line-of-sight (LoS) and keep the same function of a single curved guide. 

Many studies in the literature explore different combinations of straight and curved guides that exclude LoS, however, there is no proper approach that considers a guide system minimally excluding LoS \cite{Mildner2008,Mildner1990,Copley1993,Copley1995}. According to some authors, sometimes an extra curved guide length is necessary to guarantee unwanted neutrons and radiation \cite{Copley1993}, nevertheless, here we explore optimized scenarios. Therefore, straight-curved and curved-curved guide configurations to build the minimal S-shaped guide that excludes LoS have been investigated.

According to the literature, there is a minimum curved guide length that ensures every transported neutron touching the inner coat at least once. However, when complex guide systems using straight and curved guides are built, this minimum length is not obtained directly, and even though such geometry is simple, we consider that this approach is not well described in the correspondent bibliography.

Despite the fact that S-shaped guides are first and foremost a specific guide arrangement that provides horizontal or vertical displacement from the entrance and exit guide system, the wavelength cutoff process is the most intriguing property of such a guide \cite{DeOliveira2021,deOliveira2020,Gilles2006}. The appearance of the wavelength cutoff on final flux spectra is not found in any S-shaped model. Such a cutoff depends on curved guide characteristics, which are generally related to their length. However, as it may be seen in this paper, guide length is more important to guarantee available displacement of neutron delivery planes than to cutoff part of the energetic neutrons from the initial spectra.

Here, we have developed a study of an S-shaped guide based on curved guide arcs, which are directly bound to their characteristic angles. In addition, those guides employing LoS of whole systems have been structured to exclude epithermal neutrons and gamma rays with a minimum coating and guide walls.

There are some analyses of the so-called short guides in the literature that do not exclude LoS by their means, but with connections to other straight guides as well as curved ones \cite{Copley1995}.  With this goal, we propose the construction of the shortest guide system that guarantees LoS exclusion and relates it to the wavelength cutoff. Notwithstanding, the S-shaped guide flux performance based on the super mirror coat was checked. It has also been developed a proper formalism to build these models based on geometry coming from curved-straight and curved-curved guides.

The formalism of the Acceptance Diagram (AD) and the guidelines for neutron transport by curved direction prescribe different arrangements of super mirror indexes, i.e., from the concave and convex inner surface, with the same neutron flux for a specific wavelength range \cite{Cook2009}. Here, it has been applied the same mechanism to the construction of S-shaped guides and to investigate the relation between wavelength cutoff and neutron transport efficiency guide.

The results of this study could be useful in facilities where instruments are allocated far from the neutron source, such as the European Spallation Source (ESS) \cite{Andersen2020}. According to Zendler and Bentley, a considerable part of the 22 designed instruments is between 75 and 100 m away from the source \cite{Zendler2016}. In this scenario, neutron guide systems, which are composed of many guide sections, might exhibit a transport efficiency decrease due to misaligning sections \cite{Zendler2016,Husgard2020}. In addition, it is expected that the ESS basement foundation may present a gradual sink with time, which could systematically misalign system guide axes. In this sense, large vertical deformations are foreseen with a potential to interfere even with the neutron guides of near-source instruments. Thus, our model analysis may be applied to guide systems present in Figure 4 of Husgard's work, for instance \cite{Husgard2020}. Besides, cold neutron transportation is positively affected by the wavelength cutoff of most energetic neutrons in small-angle neutron scattering, e.g., at LoKI instrument \cite{Andersen2020}.


\section{S-Shaped Guide}
\label{SSG}
The standard guide system that we have used to describe the S-shaped guide consists of four connected guide sections, where the primary and the last are straight and the second and third guides are curved and connected in the opposite sense. Thus, we classify guide variables with the letters $p$ and $s$ referring to primary and secondary, respectively, and with numbers $2$ and $3$ to refer to second and third guides, respectively.

\begin{figure}[hbt!]
	\centering
		\includegraphics[width=0.5\textwidth]{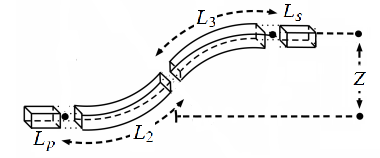}
	\caption{Side-view sketch of the S-shaped guide system. The arrangement is, sequentially, composed of a primary guide of length $L_P$, two curved guides connected in the opposite sense and with a length and curvature of $L_2$, $L_3$ and $\rho_2$, $\rho_3$, respectively, plus a secondary guide of length $L_S$. The S-shaped guide system provides a vertical (or horizontal) displacement of $Z$.}
\label{S_guide}
\end{figure}

Figure \ref{S_guide} contains the representation of an S-shaped guide system that we use as a basis of the present study. Here, variables $L_p$, $L_2$, $L_3$, and $L_s$ stand for individual guide section length, and $Z$ is the vertical (or horizontal) displacement between the entrance and exit of the guide system. 

The construction of any curved guide that excludes LoS passes, necessarily, by definition of a characteristic angle, which consequently describes the characteristic length of such guide. This angle is given by
\begin{ceqn}
\begin{align}
    \Psi_c=\sqrt{\frac{2W}{\rho}},
    \label{characteristic_ange}
\end{align}
\end{ceqn}
where $W$\footnote[1]{Here, we analyze an S-shaped guide system that can be used to provide vertical as much as horizontal displacement. The variable $W$, which corresponds to the guide width, indicates the length dimension of the guide cross-section on the same plan of curvature. So, considering a vertical displacement would make us refer to this parameter as height, but for simplicity and literature tradition sake, we maintain just $W$.} and $\rho$ stand for guide width and curvature, respectively. It is worth noting that guide width is much smaller than its curvature in a way that the system can be analyzed bidimensionally and guide height is not important in this study.

\begin{figure}[h]
	\centering
		\includegraphics[width=0.35\textwidth]{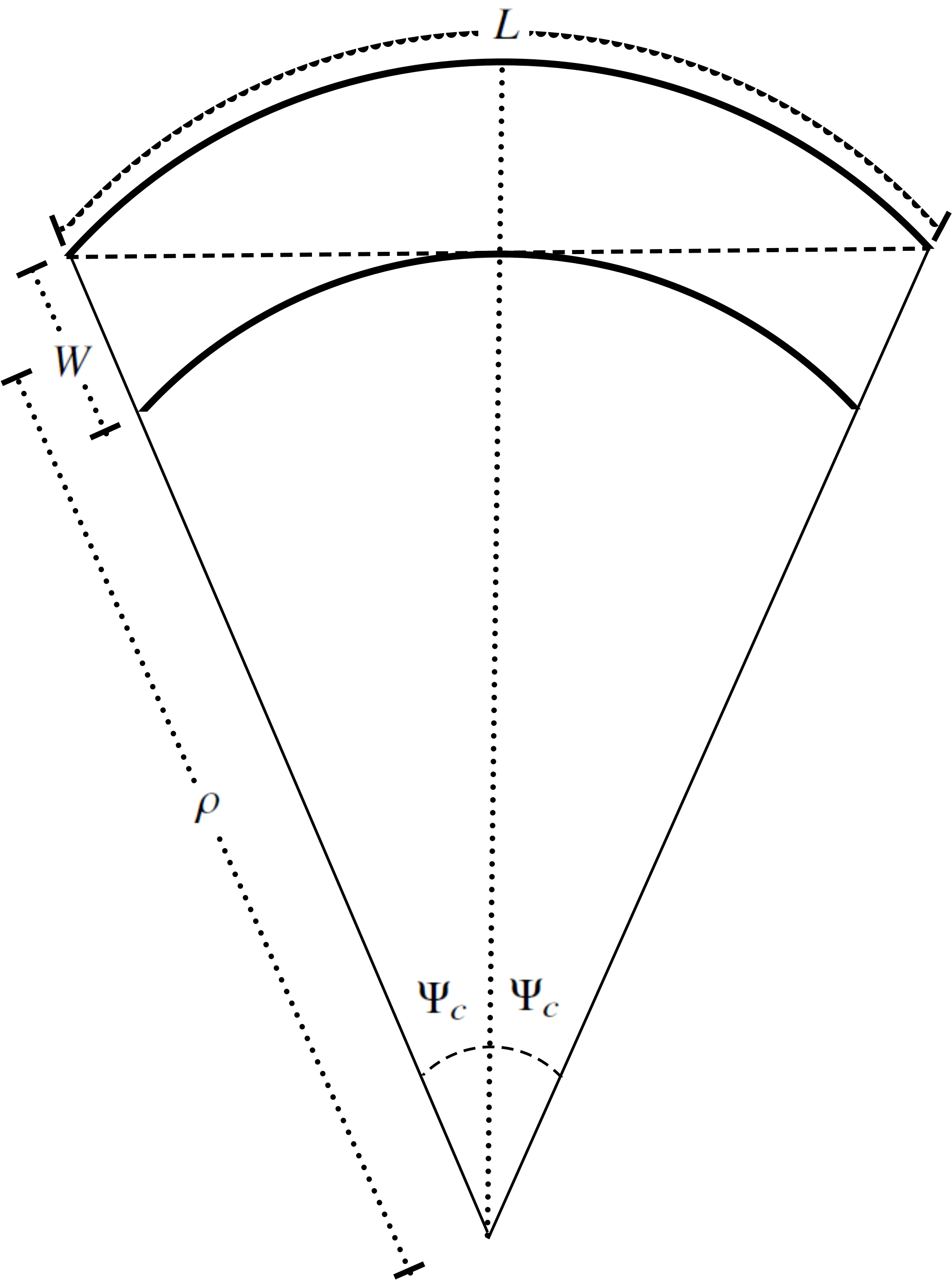}
	\caption{A curved guide with a minimum length that allows for Line-of-Sight exclusion, sketched from the side. Guide length, curvature, and width are represented by variables $L$, $\rho$, and $W$, respectively. The arc of the curved guide that excludes Line-of-Sight is given by $2\Psi_C$, where the $\Psi_C$ represents the characteristic guide angle.}
\label{LoS}
\end{figure}

In this scenario and according to Figure \ref{LoS}, a curved guide that possesses an arc of $2\Psi_c$ has a length $L_c$ and excludes LoS. Its length is written as
\begin{ceqn}
\begin{align}    
    L_c=2\Psi_c\rho=\sqrt{8W\rho}.
    \label{characteristic_length}
\end{align}
\end{ceqn}

Since neutron guides were developed to carry neutrons away from the reactor face and allow more instrument installation, their inner coating is designed to reflect incident neutrons along their whole length. Critical angles are given by

\begin{ceqn}
\begin{align}    
    \theta_c=1.73\times 10^{-3}m\lambda,
    \label{critic_angle}
\end{align}
\end{ceqn}
where any neutron of $\lambda$ with the wavelength and incident angle less than $\theta_c$ is reflected \cite{romain2016,Cook2009}. 

From the definition of critical angle and according to the literature, we define a fundamental parameter that combines aspects of both the inner coating and geometry and its known characteristic wavelength \cite{romain2016,Cook2009}. This parameter, which is also a base for defining neutron flux efficiency in curved guides, is given by
\begin{ceqn}
\begin{align}    
    \lambda_c=\frac{1}{1.73\times 10^{-3}m}\sqrt{\frac{2W}{\rho}}
    \label{characteristic_wavelength}
\end{align}
\end{ceqn}

According to the milestone work of Mildner \cite{Mildner1990}, there is a special formalism to study curved guides and to achieve AD equations. Such an approach depends on describing the neutron path utilizing two points of phase space, which are defined as the neutron radial position and inclination next to the tangential direction. These ordered pairs written as $(\Psi, z)$ and $(\Psi^{\prime},z^{\prime})$, are bound through Equation \ref{parabola_equation}\cite{Mildner1990}. 

\begin{ceqn}
\begin{align}    
    \Psi^2-{\Psi^{\prime}}^{2}=\frac{{\Psi_c}^2}{W}\left(z-z^{\prime}\right).
    \label{parabola_equation}
\end{align}
\end{ceqn}

The equation that represents the neutron trajectory, which stands for the LoS in Figure \ref{LoS}, may be achieved by substituting $(\Psi^\prime, z^\prime)$ for $(\Psi_c, W/2)$ or $(0, -W/2)$. Such equation, which has also been developed by Mildner \cite{Mildner1990}, is given by

\begin{ceqn}
\begin{align}    
    {\Psi^{2}}=\frac{{\Psi_c}^2}{W}\left(z+\frac{W}{2}\right).
    \label{LoS_main_eq}
\end{align}
\end{ceqn}
By using Equation \ref{LoS_main_eq}, different expressions that guarantee the LoS exclusion for straight-curved and curved-curved (S-shaped) guide connections have been derived. In the former case, two scenarios that allow avoiding LoS were observed. They depend on the curved section arc, where angle values may be larger or smaller than the characteristic angle. It is worthwhile noting that arcs with the same value of characteristic angle require an infinite straight guide to exclude LoS.

\begin{figure}[hbt!]
	\centering
		\includegraphics[width=0.4\textwidth]{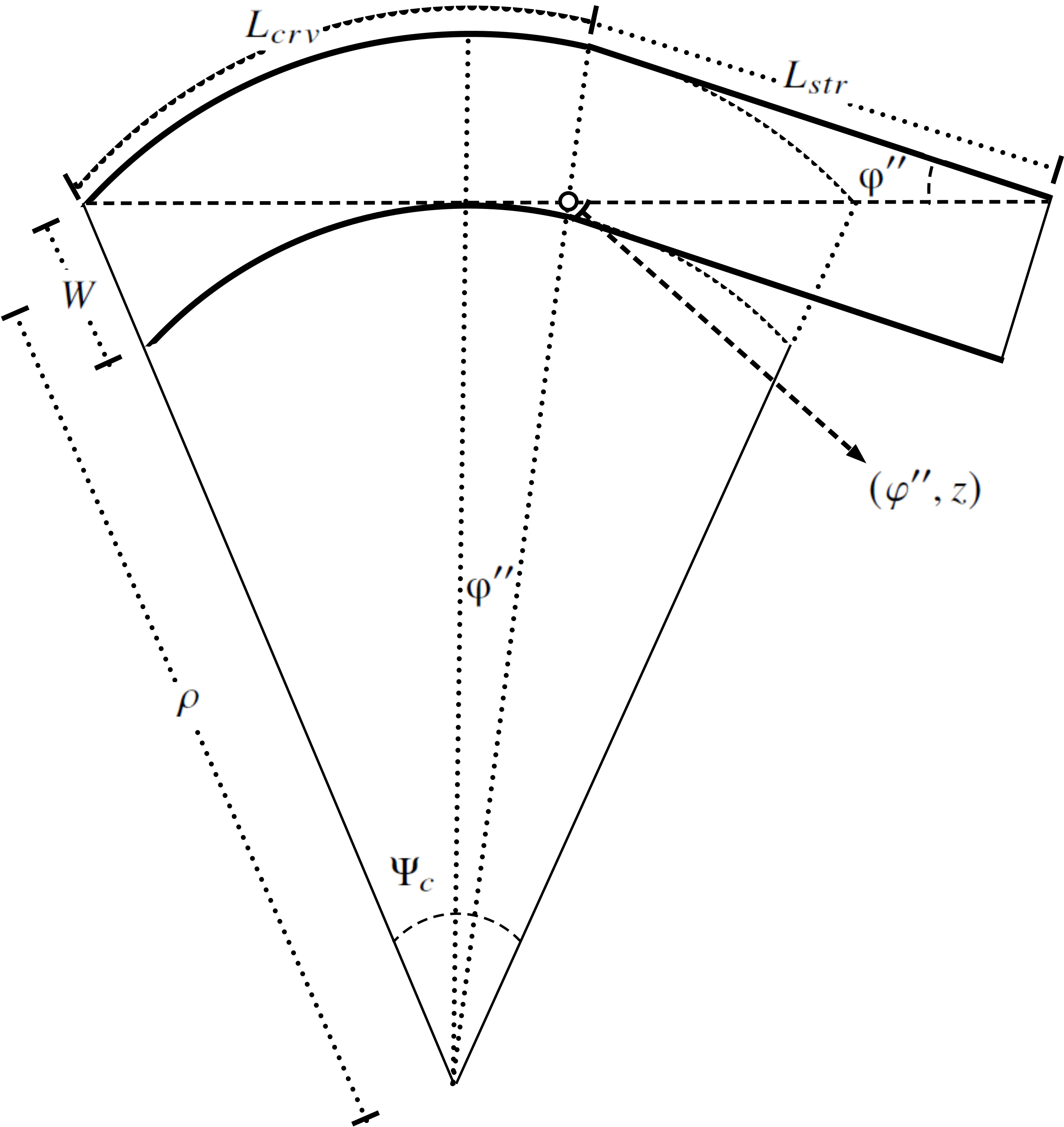}
	\caption{Side-view sketch of a curved-straight guide system connection that excludes Line-of-Sight. The straight guide, according to Figure \ref{LoS}, is attached to the right part of the curved guide, farther from the left edge. Here, the curved guide possesses an arc larger than $\Psi_C$, written as $\Psi_C+\varphi^{\prime\prime}$. Variables $\rho$ and $L_{crv}$ stand for curved guides curvature and length, respectively, $L_{str}$ stands for straight guide length, $W$ represents both guide widths and $\Psi_C$, the characteristic angle of the curved guide. The coordinate point $(\varphi^{\prime\prime},z)$ is a space-phase representation of the Acceptance Diagram formalism.}
\label{LoS_CR_STR}
\end{figure}

\begin{figure}[hbt!]
	\centering
		\includegraphics[width=0.35\textwidth]{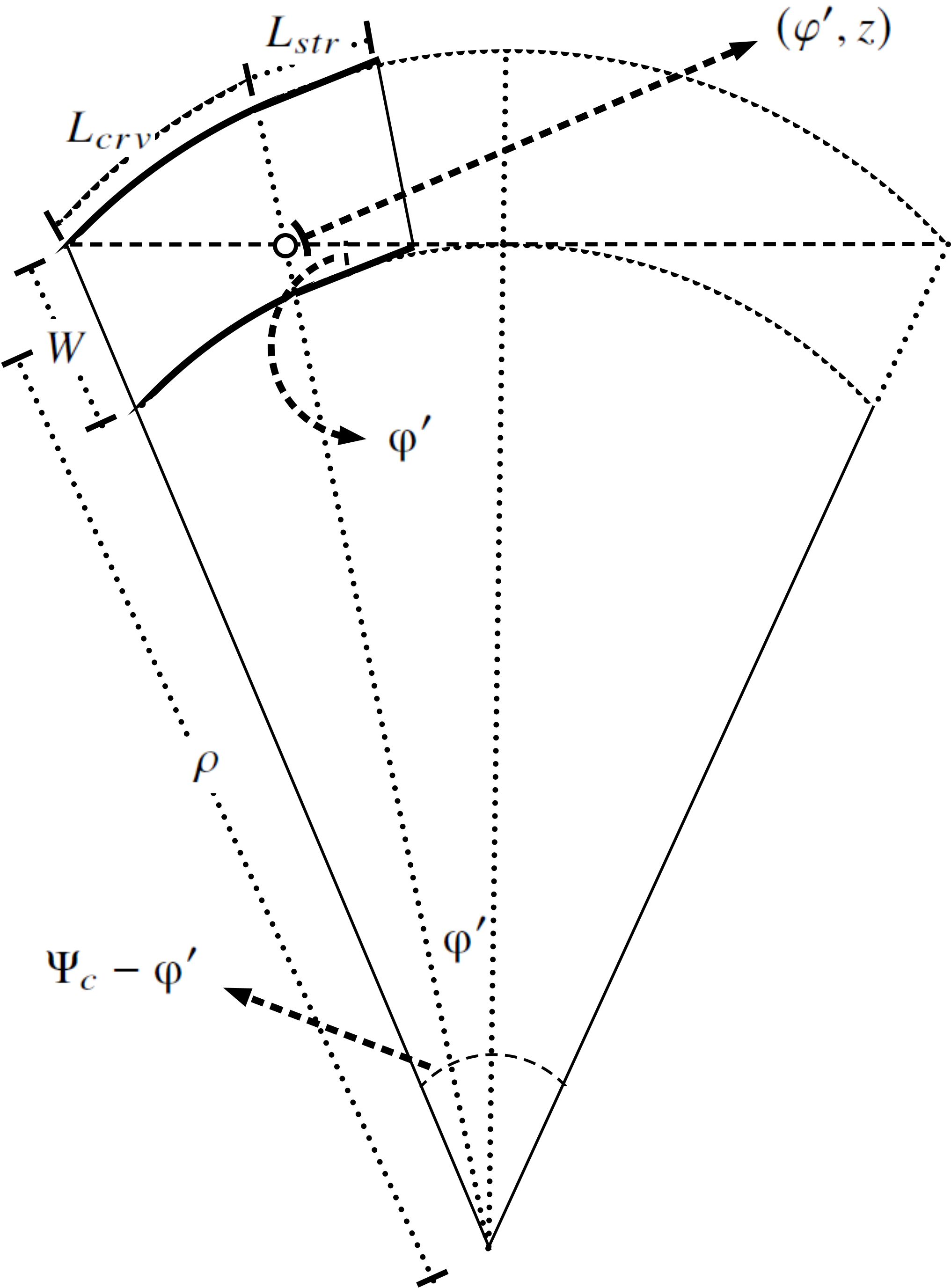}
	\caption{Side-view sketch of a curved-straight guide system connection that excludes Line-of-Sight. The straight guide, according to Figure \ref{LoS}, is attached to the left part of the curved guide, closer to the left edge. Here, the curved guide possesses an arc smaller than $\Psi_C$, written as $\Psi_C-\varphi^{\prime}$. Variables $\rho$ and $L_{crv}$ stand for curved guides curvature and length, respectively, $L_{str}$ stands for straight guide length, $W$ represents both guide widths and $\Psi_C$, the characteristic angle of the curved guide. The coordinate point $(\varphi^{\prime},z)$ is a space-phase representation of Acceptance Diagram formalism.}
\label{LoS_CR_STR_2}
\end{figure}

Straight-curved guide systems with arcs larger than $\Psi_{c2}$ are described in Figure \ref{LoS_CR_STR}, while systems with shorter arcs are presented in Figure \ref{LoS_CR_STR_2}. Variables $L_{crv}$ and $L_{str}$ represent, respectively, the length of straight and curved guide sections.

On the other hand, angles $\varphi^\prime$ and $\varphi^{\prime\prime}$ stand for main variables which describe both scenarios. Such parameters were obtained from the combination of Equation \ref{LoS_main_eq} and tangent equations derived from virtual triangles, shown in Figures \ref{LoS_CR_STR} and \ref{LoS_CR_STR_2}. Both triangles are formed by vertices given by points $(\varphi^{\prime}, z)$ and $(\varphi^{\prime}, z)$, which also compose opposite catheti in the guide radial direction, and the hypotenuses that are segments of neutron trajectories.

Since $\rho>>W$ one can approximate triangle tangents $\tan{\varphi^{\prime}}$ and $\tan{\varphi^{\prime}}$ to $\varphi^{\prime}$ and $\varphi^{\prime}$, which allows us to write the correspondent tangent equations as

\begin{ceqn}
\begin{align}    
    \varphi^{\prime(\prime\prime)}=\frac{\frac{W}{2} \pm z}{L_{str}},
    \label{tan_varphi_eq}
\end{align}
\end{ceqn}
where the single prime ($^{\prime}$) angle stands for the plus ($+$) sign and the double prime angle ($^{\prime\prime}$) for the minus sign ($-$). In these terms, we substitute $(\Psi, z)$ by $( \varphi^{\prime}, z)$ and $( \varphi^{\prime\prime}, z)$ in Equation \ref{LoS_main_eq} and combine results, respectively, to Equation \ref{tan_varphi_eq} variants to eliminate the $z$ variable. After doing this process, we obtain Equations \ref{varphi_line} and \ref{varphi_lineline}, which are given by

\begin{ceqn}
\begin{align}    
    \varphi^{\prime}=\frac{2L_{str}}{\rho},
    \label{varphi_line}
\end{align}
\end{ceqn}

\begin{ceqn}
\begin{align}    
    \varphi^{\prime\prime}=\sqrt{{\Psi_c}^2+\frac{L_{str}^2}{\rho^2}}-\frac{L_{str}}{\rho}.
    \label{varphi_lineline}    
\end{align}
\end{ceqn}
From these two angles, i.e., $\varphi^{\prime}$ and $\varphi^{\prime\prime}$, we can define curved guide arcs, which are preceded or followed by a straight guide and preserve LoS exclusion. Once S-shaped guide edges are characterized, how the curved-curved guide connection affects those arc angles can be analyzed.

\begin{figure}[hbt!]
	\centering
		\includegraphics[width=0.4\textwidth]{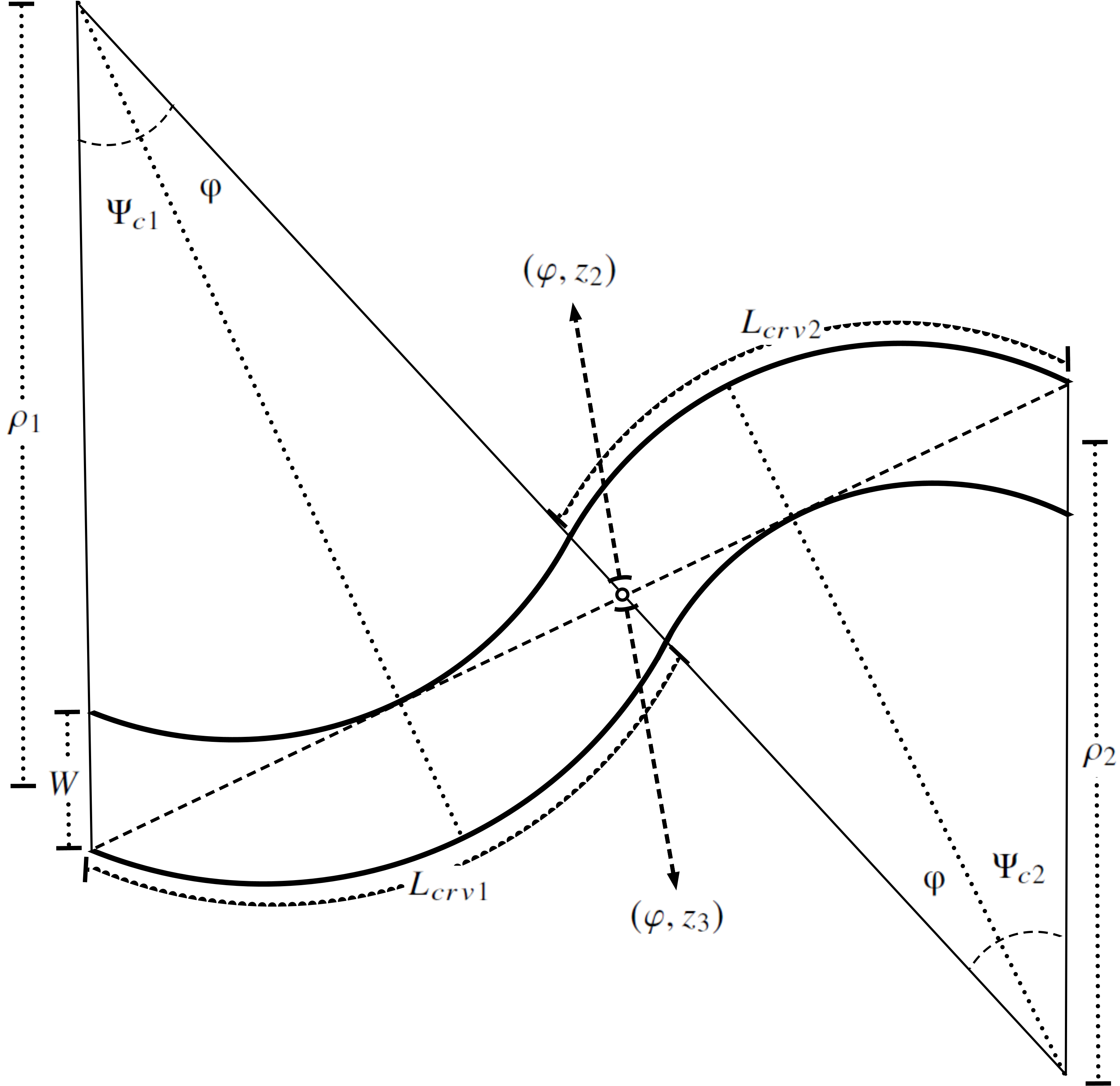}
	\caption{Side-view sketch of a curved-curved guide system connection in the opposite sense and that excludes Line-of-Sight. The curvatures and lengths of curved guides are given by $\rho_1$ and $\rho_2$, and $L_{str1}$ and $L_{str2}$, respectively and sequentially from left to right. Characteristic angles $\Psi_{C1}$ and $\Psi_{C2}$, and coordinates points $(\varphi,z_1)$ and $(\varphi,z_2)$ of Acceptance Diagram formalism, also, correspond sequentially to each curved guide as much as arcs that are written as $\Psi_{C1}+\varphi$ and $\Psi_{C2}+\varphi$. The angle $\varphi$ represents a rotation of curved guides that make their trajectories, as presented in Figure \ref{LoS}, coincide.}
\label{LoS_CR_CR}
\end{figure}

In Figure \ref{LoS_CR_CR}, we observe a sketch that represents an S-shaped guide curved connection. In this scenario, there is a mutual angle $\varphi$ that represents the same rotation on curved guides and, consequently, the inclination of neutron trajectory next to both guides reference system. According to geometry, such an angle composes both the angular parts of the ordered pair shown in Figure \ref{LoS_CR_CR}. Here this point, which represents a single point on each curved guide reference, is given as $  (\varphi, z_{2})$ for the first curved guide and  $(\varphi, z_{3})$ for the second one.

By substituting two points on Equation \ref{LoS_main_eq}, we obtain a pair of equations, namely Equations \ref{varphi_2} and \ref{varphi_3}, representing neutron trajectory according to two curved guides.

\begin{ceqn}
\begin{align}    
    \varphi^{2}=\frac{{\Psi_{c2}}^2}{W}\left(z_2+\frac{W}{2}\right)
    \label{varphi_2}
\end{align}
\end{ceqn}

\begin{ceqn}
\begin{align}    
    \varphi^{2}=\frac{{\Psi_{c3}}^2}{W}\left(z_3+\frac{W}{2}\right)
    \label{varphi_3}
\end{align}
\end{ceqn}

 Notwithstanding, and aiming to define the $\varphi$ equation, we are able to correlate variables $z_2$ and $z_3$ by noticing that these values on the interface between guides may be written as $z_2=-z_3$. By means of this relation, Equations \ref{varphi_2} and \ref{varphi_3} can be combined and $\varphi$ then derived, which is given by

\begin{figure}[hbt!]
	\centering
		\includegraphics[width=0.4\textwidth]{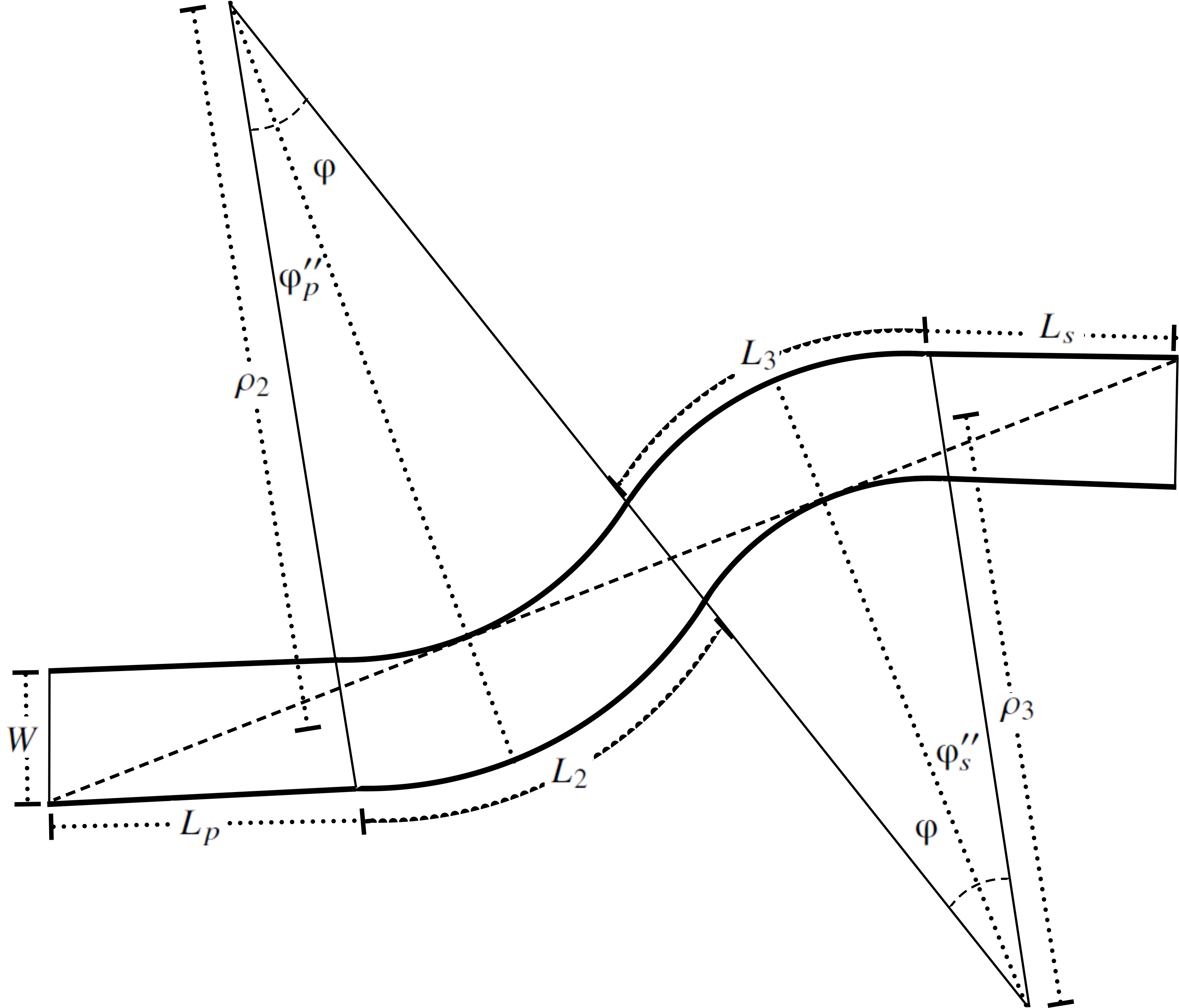}
	\caption{Side-view sketch of the S-shaped guide system. The arrangement is sequentially composed of a primary guide of length $L_P$, two curved guides connected in the opposite sense and with a length and curvature of $L_2$, $L_3$ and $\rho_2$, $\rho_3$, respectively, and a secondary guide of length $L_S$. The arcs of curved guides are given sequentially by $\varphi+\varphi_{P}^{\prime\prime}$ and $\varphi+\varphi_{S}^{\prime\prime}$. The variable $W$ stands for guide system width.}
\label{LoS_STR_CR_CR_STR}
\end{figure}

\begin{ceqn}
\begin{align}    
    \varphi=\sqrt{\frac{2W}{\rho_2+\rho_3}}.
    \label{varphi}
\end{align}
\end{ceqn}
After defining angles $ \varphi$, $\varphi^{\prime}$ and $\varphi^{\prime\prime}$, described by Equations \ref{varphi}, \ref{varphi_line} and \ref{varphi_lineline} respectively, we can characterize the entire S-shaped guide system that guarantees LoS exclusion. This complete system is shown in Figure \ref{LoS_STR_CR_CR_STR} with their correspondent angles addressed on  Equations \ref{varphi_p_line} and \ref{varphi_s_line}.

It is worth noting that Figure \ref{LoS_STR_CR_CR_STR} is just one of the configurations that exclude LoS. Other scenarios can be obtained by using straight-curved guide combinations of Figure \ref{LoS_CR_STR_2} at one of the S-shaped guide edges. Despite all cases have been investigated, just the ``classical'' S-shaped guide sketch has been presented to simplify. Here we investigate if all these configurations may be called a proper S-shaped guide, i.e., if they impose wavelength cutoff on the transported neutron profile.

\begin{ceqn}
\begin{align}    
    \varphi_p^{\prime}=\frac{2L_p}{\rho_2}, \quad     \varphi_p^{\prime\prime}=\sqrt{{\Psi_{c2}}^2+\frac{L_1^2}{\rho_2^2}}-\frac{L_1}{\rho_2},
\label{varphi_p_line}    
\end{align}
\end{ceqn}

\begin{ceqn}
\begin{align}    
    \varphi_s^{\prime}=\frac{2L_s}{\rho_3}, \quad     \varphi_s^{\prime\prime}=\sqrt{{\Psi_{c3}}^2+\frac{L_4^2}{\rho_3^2}}-\frac{L_4}{\rho_3}.
\label{varphi_s_line}    
\end{align}
\end{ceqn}

On the basis of these equations, the three scenarios of S-shaped guides according to their curved guide arcs may be defined, where $\gamma_2$ and $\gamma_3$ stand, respectively, at their values. So, the curved guide system is described by
\textbf{
\begin{enumerate}[a.]
\item\label{a} $\gamma_2=\varphi-\varphi_p^{\prime},\quad \gamma_3=\varphi + \varphi_s^{\prime\prime}$
\item\label{b} $ \gamma_2=\varphi + \varphi_p^{\prime\prime},\quad \gamma_3=\varphi- \varphi_s^{\prime}$
\item\label{c} $ \gamma_2=\varphi + \varphi_p^{\prime\prime},\quad \gamma_3 =\varphi + \varphi_s^{\prime\prime}$
\end{enumerate}}
From now on, by this letter classification for identifying simulation cases, where case \textbf{c.} corresponds to, as already mentioned, the ``classical'' S-shaped guide of Figure \ref{LoS_STR_CR_CR_STR}.

Maybe the unique characteristic of an S-shaped guide corresponds to the wavelength cutoff that its geometry imposes on neutron flux spectra. According to AD formalism, there is a theoretical cutoff value, which is proportional to the characteristic wavelength, written as $\frac{\lambda_c}{\sqrt{2}}$ and geometrically indicates a cutoff angle of $\frac{\Psi_c}{\sqrt{2}}$ \cite{Mildner1990,deOliveira2020}. Such a result is coherent with Equation \ref{varphi} when both curved guides possess the same curvature, i.e., when $\rho_2=\rho_3=\rho$ we get $\varphi = \sqrt{\frac{W}{\rho}}=\frac{\Psi_c}{\sqrt{2}}$.  Under these conditions, one can write the wavelength cutoff of a ``classical'' S-shaped guide (configuration \textbf{c.}) as
\begin{ceqn}
\begin{align}    
    \lambda_{cut}=\frac{1}{1.73\times 10^{-3}m}\varphi.
\label{lambda_cut}    
\end{align}
\end{ceqn}

Checking the validity of such an expression is one of the scopes present in this work, as much as verifying the conditions of the wavelength cutoff for configurations \textbf{a.} and \textbf{b.}. Another important point to verify is the relation between the wavelength cutoff and the super mirror coating indexes m. In other words, we intend to verify if the same formalism of curved guide inner indexes may be assumed for the S-shaped guides. There is, in the literature, a well-established process dictating that different concave and convex surface mirror indexes allow us, depending on their values, to maintain the same neutron flux in cases of equal indexes \cite{Cook2009}. 

 Here we define a variable $m_{out}$ as being concave index and variable $m_{in}$ as the convex index surface. The basis of estimating neutron transport by a curved guide is, according to the literature, the characteristic wavelength \cite{romain2016,deOliveira2020,Cook2009}, which is written as  

\begin{ceqn}
\begin{align}    
    \lambda_{c}=\frac{1}{1.73\times 10^{-3}m_{out}}\Psi_c.
\label{lambda_c}    
\end{align}
\end{ceqn}

This variable, which represents a hybrid combination of guide geometry and inner coat index values, imposes that the neutrons with lower and higher wavelength values be transmitted by curved guides with efficiencies lower and higher than $66.67$~$\%$. Such information comes from AD formalism and neutron reflection regimes, which can be Garland and Zigzag. The former regime corresponds to neutron reflections only on the concave surface, the latter, on the other hand, indicates neutron reflection on both surfaces. Garland regime represents a neutron transport with efficiency less than $66.67$~$\%$ \cite{romain2016,Mildner1990,Copley1993}.

According to the literature, a super mirror index relation of $m_{out}<m_{in}$ guarantees a wavelength gap, where neutron transport is the same as a guide with $m_{out}=m_{in}$ \cite{Cook2009}. Such range is given by neutrons with wavelength $\lambda$ that satisfy the relation $\lambda_c<\lambda<\lambda^{\prime}$, where the upper variable value is written as 

\begin{ceqn}
\begin{align}    
    \lambda^{\prime}=\frac{m_{out}}{\sqrt{m_{out}^2-m_{in}^2}}\lambda_c.
\label{lambda_prime}    
\end{align}
\end{ceqn}

In these terms, we propose an investigation of different index coats next to the wavelength cutoff and the neutron transport efficiency according to three typical reactor sources, i.e., cold, thermal, and hot sources \cite{sourcegen}. The use of convex indexes with lower values other than concave, allows savings on system guide building and it, also, guarantees the tailoring process phase, which provides a homogeneous neutron distribution of overall guide divergences \cite{Mildner2008}.

Here, there is an important detail that it has to be taken into account. Equation \ref{varphi}, and consequently Equation \ref{lambda_cut}, are deduced by considering that all surface coats possess the same index values. Therefore, we would not be able to investigate cases where curved guides of the S-shaped guide system have different index values. We propose a solution based on the $\lambda_c$ central role in neutron transportation efficiency. Since $\lambda_c$ settles the efficiency value of $66.67$~$\%$, it is possible to keep both guide concave indexes with the same value and compensate the characteristic wavelength value by changing the curvature value. Of course, this change is purely virtual, but with it, we are still able to use Equations \ref{varphi} and \ref{lambda_cut} to estimate the wavelength cutoff. Therefore, to keep $\lambda_c$ value it is necessary to follow the relation $\rho m^2=\rho^* m^{*2}$, which can be rewritten into
\begin{ceqn}
\begin{align}    
    \rho_{i}^{*}=\frac{m_{out,i}^2}{m_{out,i}^{* 2}}\rho_{i},
\end{align}
\end{ceqn}
for $i=2$ or $i=3$ and where the symbol $^{*}$ stands for new variable values. Here we emphasize that this change has to be applied just on one of the curved guides to make both $m_{out}$ values equal, i.e., $m_{out2}=m_{out3}^{*}$ for keeping $m_{out2}$ and $m_{out2}^{*}=m_{out3}$ for a constant $m_{out3}$ value. Therefore, a rewritten version of Equation \ref{varphi}, corresponding to the former case, could be given by

\begin{ceqn}
\begin{align}    
    \varphi^{*}=\sqrt{\frac{2W}{\rho_{2}+\rho_{3}^{*}}},
\label{new_varphi}    
\end{align}
\end{ceqn}
where $ \varphi^{*}$ represents the new value of $\varphi$, giving rise to a new wavelength cutoff expression, described as 

\begin{ceqn}
\begin{align}    
    \lambda_{cut}^{*}=\frac{1}{1.73\times 10^{-3}m}\varphi^{*},
\label{new_lambda_cut}    
\end{align}
\end{ceqn}
where $m=m_{out2}=m_{out3}^{*}$.

The last important point to be mentioned is the vertical (or horizontal) displacement that S-shaped guides provide. According to de Oliveira \cite{deOliveira2020}, we observe that the system present in Figure \ref{LoS_STR_CR_CR_STR} possess a $Z$ displacement given by

\begin{ceqn}
\begin{align}    
    Z=\rho_2\left[1-\cos \left(\frac{L_2}{\rho_2}\right)\right]+\rho_3\left[1-\cos \left(\frac{L_3}{\rho_3}\right)\right]
    \label{Z}
\end{align}
\end{ceqn}

Since one of our goals is based on studying S-shaped guide wavelength cutoff properties, we have to compare scenarios of curved guides with different lengths next to cases with $\gamma_2=2\Psi_2$ and $\gamma_3=2\Psi_3$. As previously seen in Figure \ref{LoS_STR_CR_CR_STR}, minimal scenarios that exclude LoS have necessarily lower arc values than $2\Psi_2$ and $2\Psi_3$. By considering this fact, we observe that cases with $\gamma_2<2\Psi_2$ and $\gamma_3<2\Psi_3$, according to Equation \ref{Z} may never have displacements $Z$ higher than $8W$, obtained by substituting $L_2 = \sqrt{8W\rho_2}$ and $L_3 = \sqrt{8W\rho_3}$ on Equations \ref{Z} and assuming that $L_2<<\rho_2$ and $L_3<<\rho_3$.

Considering the curved guide arcs of the S-shaped guide system crucial for investigating its properties, we define the variable R that represents the ratio of the angular arc of both curved guides by two times their corresponding characteristic angle, in agreement with the minimal arc that individually excludes LoS. Such a ratio is given by

\begin{ceqn}
\begin{align}    
    R_{2/3} = \frac{\gamma_i}{2\Psi_{c2/3}},
\label{R_2_3}    
\end{align}
\end{ceqn}
where the lower indexes 2 and 3 stand, respectively, for the first and second curved guide of the system. In Figure \ref{LoS_STR_CR_CR_STR} it is possible to observe that curved guide arcs are important for defining primary and secondary guide lengths, while the minimum LoS criterion is kept. According to the literature, the length of a straight guide located after a curved one is important to keep neutron flux, homogeneously, distributed at the end of the guide system \cite{deOliveira2020}.

In addition, the length of all guides is also important to define an S-shaped guide system project, since available space in nuclear facilities may be a crucial task. In this scenario, Equations \ref{varphi} and \ref{R_2_3} are used to write a straight guide length in terms of other variables. This expression, which is defined according to configuration \textbf{c.}, is written as
\begin{ceqn}
\begin{align}    
    L_{p/s} = \frac{\rho_{2/3}}{2}\left[\frac{\Psi^2_{c2/3}-\left(2\Psi_{c2/3}R_{2/3}-\varphi\right)^2}{\left(2\Psi_{c2/3}R_{2/3}-\varphi\right)}\right]
\label{L_P_S}    
\end{align}
\end{ceqn}

As we can see in Equation \ref{L_P_S}, there is a singularity imposed by the divisor that is composed of two classes of variables, individual guide curvatures $\rho_2$ and $\rho_3$, from characteristic angle $\Psi_{C2/3}$, and mutual curved guide curvature from parameter $\varphi$. When both curvatures are different, the singularity occurs for $R_{2/3}=\frac{1}{2}\sqrt{\frac{\rho_{2/3}}{\rho_2+\rho_3}}$, otherwise for $\rho_2=\rho_3$ we have a fix singularity at ratio $R_{2/3}=35.35$~$\%$.

\section{MCSTAS Simulations}
\label{MS}
In this section, we present relevant information on MCSTAS simulations performed. We intend to investigate three different aspects of the proposed S-shaped guide system, and each case has a corresponding series of simulations.

The very first sequence of simulations is carried out to check the LoS aspects of the wavelength cutoff properties. Here, configurations \textbf{a.}, \textbf{b.}, and \textbf{c.} are tested. The second sequence of simulations consists of exploring Equation \ref{L_P_S} primary and secondary straight guide lengths, intended to define a minimal arc value that guarantees a wavelength cutoff. The third series of simulations consists of exploring configuration \textbf{c.} in scenarios of different concave and convex surface indexes ($m_{out}>m_{in}$). Simulations are performed by taking two cases from the first sequence as the basis. Considering these results, we intend to validate Equation \ref{new_lambda_cut}, which applies a sort of reset on m indexes by equalizing their values through geometrical changing parameters. In addition, and according to these results, we aim at compact S-shaped guides with lower cost, similar properties, and compatible neutron transportation efficiency (similar to the simple curved guide design formalism) \cite{Cook2009}.

\begin{figure}[hbt!]
	\centering
		\includegraphics[width=0.7\textwidth]{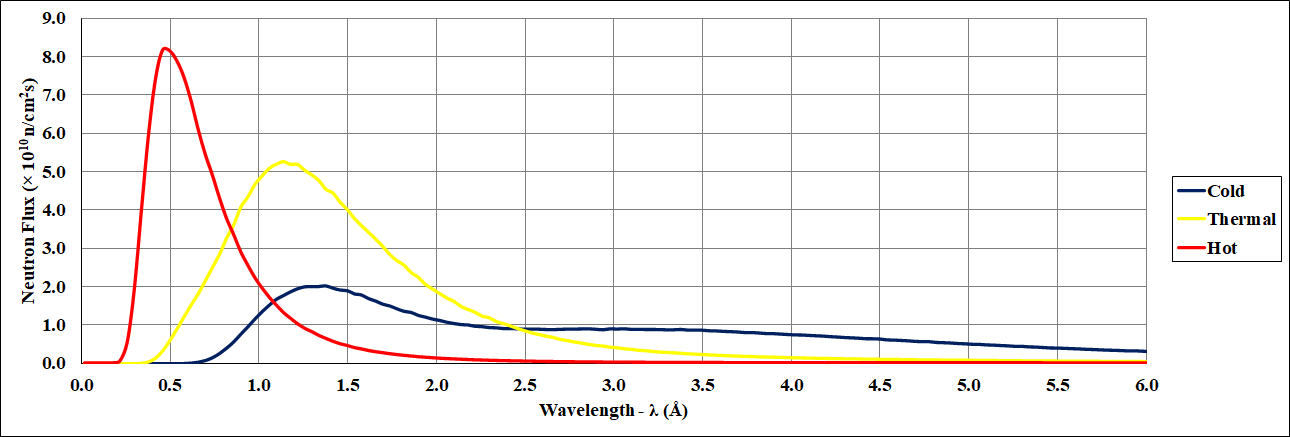}
	\caption{Virtual source profiles of the ILL sources from the MCSTAS instrument component. Red, yellow and blue curves stand for ILL hot, thermal, and cold sources, respectively.}
\label{SOURCE}
\end{figure}

Simulations are carried out through MCSTAS 3.0 software and neutron sources are defined using the tool \verb|Source_gen()|, which allows us to mimic different wavelength neutron distributions based on Maxwellian distributions. For simulations, we have picked Maxwellian parameters that correspond to three types of \textit{Institut Laue-Langevin} - ILL sources. Such sources correspond to cold, thermal, and hot neutron wavelength specter profiles \cite{sourcegen}. We use them to test S-shaped guides applicability to different scenarios according to their performance in neutron transportation. The wavelength profiles of three ILL sources are presented in Figure \ref{SOURCE}.

Simulation cases depend on geometrical and coating surface parameters. The former is composed of straight guide lengths, $L_{P}$ and $L_{S}$ and curved guide curvatures, $\rho_2$ and $\rho_3$. Here, width and height guides are fixed every $5$~$cm$. On the other hand, coating variables are given by $m_1$ and $m_4$ as indexes of all primary and secondary straight guide surfaces, respectively, $m_{out2}$ and $m_{out3}$ as concave curved guide surfaces, and $m_{in2}$ and $m_{in3}$ correspond to convex curved guide surfaces. In addition, we standardized that lateral curved guide surfaces possess the same mirror index as their concave surfaces, i.e., $m_{out2}$ for the first curved guide and $m_{out3}$ for the second one.

We define simulation cases with different coating indexes by following the relation,
\begin{ceqn}
\begin{align}    
    m_{1}\geq m_{out2}\geq m_{out3}\geq m_{in2}\geq m_{in3}\geq m_{4},
    \label{escala_m}
\end{align}
\end{ceqn}
which guarantees that higher indexes be always located nearer to the source than to the system exit. In addition, this relation provides a phase space tailoring process at the end of the S-shaped guide system. This procedure dictates that $m_{in3}$ should be equal to $m_4$ in order to provide a uniform divergence of neutron distribution at the instrument entrance.

The first series of simulations is composed of multiple cases described in Table \ref{tbl1}. They are made up of twelve cases with different variable combination values $L_{p}$, $L_{s}$, $\rho_{2}$, and $\rho_{3}$, where each one of them is subdivided into configuration \textbf{a.}, \textbf{b.} and \textbf{c.}. We address Roman numbers (\textbf{I} - \textbf{XII}) to identify cases and letters $^{a}$, $^{b}$, and $^{c}$ to specify, respectively, their subcases.

In addition, there are values of curved guide lengths in Table \ref{tbl1}, which are obtained by simply multiplying curved guide arcs by their curvature, i.e., $L_i=\rho_{i}\gamma_{i}$ for $i=2$ or $i=3$.

\begin{table}[hbt!]
\caption{Configurations of the first sequence of simulations. Twelve cases, namely \textbf{I} to \textbf{XII}, are subdivided into three other cases, namely \textbf{a.}, \textbf{b.}, and \textbf{c.}. The main divisions stand for geometrical parameter values, guide curvatures, and lengths, while subdivisions represent geometrical guide disposal.}\label{tbl1}
\begin{tabular*}{\tblwidth}{@{} LLLLLLL@{} }
\toprule
\textbf{Case}     & \textbf{$L_p (m)$}    & \textbf{$\rho_2 (m)$}    & \textbf{$L_2 (m)$} & \textbf{$\rho_3 (m)$}    & \textbf{$L_3 (m)$} & \textbf{$L_s (m)$}    \\
\midrule 
\textbf{I$^a$}    & \multirow{3}{*}{1.00} & \multirow{3}{*}{4000.00} & 33.17              & \multirow{3}{*}{4000.00} & 12.14              & \multirow{3}{*}{1.00} \\ 
\textbf{I$^b$}    &                       &                          & 12.14              &                          & 33.17              &                       \\  
\textbf{I$^c$}    &                       &                          & 33.17              &                          & 33.17              &                       \\ 
\textbf{II$^a$}   & \multirow{3}{*}{1.00} & \multirow{3}{*}{4000.00} & 35.35              & \multirow{3}{*}{2000.00} & 6.16               & \multirow{3}{*}{1.00} \\ 
\textbf{II$^b$}   &                       &                          & 14.33              &                          & 21.34              &                       \\ 
\textbf{II$^c$}   &                       &                          & 35.35              &                          & 21.34              &                       \\ 
\textbf{III$^a$}  & \multirow{3}{*}{1.00} & \multirow{3}{*}{4000.00} & 36.91              & \multirow{3}{*}{1000.00} & 2.47               & \multirow{3}{*}{1.00} \\ 
\textbf{III$^b$}  &                       &                          & 15.89              &                          & 13.52              &                       \\ 
\textbf{III$^c$}  &                       &                          & 36.91              &                          & 13.52              &                       \\ 
\textbf{IV$^a$}   & \multirow{3}{*}{1.00} & \multirow{3}{*}{2000.00} & 23.18              & \multirow{3}{*}{2000.00} & 8.00               & \multirow{3}{*}{1.00} \\ 
\textbf{IV$^b$}   &                       &                          & 8.00               &                          & 23.18              &                       \\ 
\textbf{IV$^c$}   &                       &                          & 23.18              &                          & 23.18              &                       \\ 
\textbf{V$^a$}    & \multirow{3}{*}{1.00} & \multirow{3}{*}{2000.00} & 24.72              & \multirow{3}{*}{1000.00} & 3.77               & \multirow{3}{*}{1.00} \\ 
\textbf{V$^b$}    &                       &                          & 9.55               &                          & 14.82              &                       \\ 
\textbf{V$^c$}    &                       &                          & 24.72              &                          & 14.82              &                       \\ 
\textbf{VI$^a$}   & \multirow{3}{*}{1.00} & \multirow{3}{*}{1000.00} & 16.12              & \multirow{3}{*}{1000.00} & 5.07               & \multirow{3}{*}{1.00} \\ 
\textbf{VI$^b$}   &                       &                          & 5.07               &                          & 16.12              &                       \\ 
\textbf{VI$^c$}   &                       &                          & 16.12              &                          & 16.12              &                       \\ 
\textbf{VII$^a$}  & \multirow{3}{*}{3.00} & \multirow{3}{*}{4000.00} & 31.37              & \multirow{3}{*}{4000.00} & 10.14              & \multirow{3}{*}{2.00} \\ 
\textbf{VII$^b$}  &                       &                          & 8.14               &                          & 32.24              &                       \\ 
\textbf{VII$^c$}  &                       &                          & 31.37              &                          & 32.24              &                       \\ 
\textbf{VIII$^a$} & \multirow{3}{*}{3.00} & \multirow{3}{*}{4000.00} & 33.55              & \multirow{3}{*}{2000.00} & 4.16               & \multirow{3}{*}{2.00} \\ 
\textbf{VIII$^b$} &                       &                          & 10.33              &                          & 20.45              &                       \\ 
\textbf{VIII$^c$} &                       &                          & 33.55              &                          & 20.45              &                       \\ 
\textbf{IX$^a$}   & \multirow{3}{*}{3.00} & \multirow{3}{*}{4000.00} & 35.11              & \multirow{3}{*}{1000.00} & 0.47               & \multirow{3}{*}{2.00} \\ 
\textbf{IX$^b$}   &                       &                          & 11.89              &                          & 12.67              &                       \\ 
\textbf{IX$^c$}   &                       &                          & 35.11              &                          & 12.67              &                       \\ 
\textbf{X$^a$}    & \multirow{3}{*}{3.00} & \multirow{3}{*}{2000.00} & 21.46              & \multirow{3}{*}{2000.00} & 6.00               & \multirow{3}{*}{2.00} \\ 
\textbf{X$^b$}    &                       &                          & 4.00               &                          & 22.28              &                       \\ 
\textbf{X$^c$}    &                       &                          & 21.46              &                          & 22.28              &                       \\ 
\textbf{XI$^a$}   & \multirow{3}{*}{3.00} & \multirow{3}{*}{2000.00} & 23.00              & \multirow{3}{*}{1000.00} & 1.77               & \multirow{3}{*}{2.00} \\ 
\textbf{XI$^b$}   &                       &                          & 5.55               &                          & 13.97              &                       \\ 
\textbf{XI$^c$}   &                       &                          & 23.00              &                          & 13.97              &                       \\ 
\textbf{XII$^a$}  & \multirow{3}{*}{3.00} & \multirow{3}{*}{1000.00} & 14.51              & \multirow{3}{*}{1000.00} & 3.07               & \multirow{3}{*}{2.00} \\ 
\textbf{XII$^b$}  &                       &                          & 1.07               &                          & 15.27              &                       \\ 
\textbf{XII$^c$}  &                       &                          & 14.51              &                          & 15.27              & \\
\bottomrule
\end{tabular*}
\vspace{-0.4cm}
\begin{flushleft}
{\footnotesize{$^a$ $\gamma_2=\varphi+\varphi_{p}^{\prime\prime}$,   $\gamma_3=\varphi-\varphi_{s}^{\prime}$}}\\

{\footnotesize{$^b$ $\gamma_2=\varphi-\varphi_{p}^{\prime}$,   $\gamma_3=\varphi+\varphi_{s}^{\prime\prime}$}}\\

{\footnotesize{$^c$ $\gamma_2=\varphi+\varphi_{p}^{\prime\prime}$,   $\gamma_3=\varphi+\varphi_{s}^{\prime\prime}$}}
\end{flushleft}
\end{table}

Cases of Table \ref{tbl1} are carried out and results are disposed of Table \ref{VACA} and compacted on the graphic of Figure \ref{R_CUTOFF}. We note here that most of the good agreement between the theoretical and simulated cutoff corresponds predominantly to configuration \textbf{c.} Such behavior is due to configurations \textbf{a.} and \textbf{b.} principles that always impose shorter arcs than the other one. The second set of simulations explores precisely this aspect because primary and secondary straight guide length variation force curved guide arc variations. Based on these results, we intend to determine minimal arc values that guarantee a wavelength cutoff according to the theoretical value. Further information on these results is presented in Section \ref{RAD}.

We investigate arc properties by picking S-shaped guide systems with equal curvature values, namely $\rho=2000$~$m$ and $\rho = 1000$~$m$. In these scenarios, we simulate arc ratios of different values, by varying from $40$~$\%$ to $80$~$\%$ in a $2$~$\%$ step sequence. These twenty-one proposed cases of simulation, whose values are based on Equation \ref{L_P_S}, are shown in Table \ref{table_L_P_L_S}.

\begin{table}[hbt!]
\caption{Configurations of the second sequence of simulations. Twenty-one cases present correspondent values of arc ratios $R_{2}$ and $R_{3}$, and straight primary and secondary guides, respectively. Simulation ratio values range from $40$~$\%$ to $80$~$\%$ in steps of $2$~$\%$ and corresponds to two different S-shaped guide systems, both with equal curvature values, i.e. $\rho_2=\rho_3$, and given at $2000$~$m$ and $1000$~$m$.}\label{table_L_P_L_S}
\begin{tabular*}{\tblwidth}{@{} LCCLCC@{} }
\toprule
\multirow{1}{*}{$R_{2/3}$} & $\rho = 2000$~$m$   & $\rho = 1000$~$m$ $^{\dagger}$ & \multirow{1}{*}{$R_{2/3}$} & $\rho = 2000$~$m$   & $\rho = 1000$~$m$ $^{\dagger}$ \\
\multirow{1}{*}{$(\%)$}      &  \multicolumn{2}{c}{$L_{P/S}(m)$} & \multirow{1}{*}{$(\%)$} & \multicolumn{2}{c}{$L_{P/S}(m)$}\\
\midrule 

$40$ & $75.46$ & $53.36$  & $62$ & $9.50$ & $6.72$ \\
$42$ & $52.27$ & $36.96$ & $64$ & $8.29$ & $5.86$ \\
$44$ & $39.68$ & $28.06$ & $66$ & $7.20$ & $5.09$ \\
$46$ & $31.71$ & $22.42$ & $68$ & $6.21$ & $4.39$ \\
$48$ & $26.17$ & $18.51$ & $70$ & $5.31$ & $3.75$ \\
$50$ & $22.7$ & $15.61$  & $72$ & $4.47$ & $3.16$ \\
$52$ & $18.89$ & $13.36$ & $74$ & $3.68$ & $2.60$ \\
$54$ & $16.33$ & $11.54$ & $76$ & $2.95$ & $2.09$ \\
$56$ & $14.21$ & $10.05$ & $78$ & $2.26$ & $1.60$ \\
$58$ & $12.41$ & $8.78$ & $80$ & $1.61$ & $1.14$ \\
$60$ & $10.86$ & $7.68$ & $-$ & $-$ & $-$\\

\bottomrule
\end{tabular*}
\vspace{-0.2cm}
\begin{flushleft}
\footnotesize{$^{\dagger}$ $\rho = \rho_2 = \rho_3$}
\end{flushleft}

\end{table}

\begin{figure}[hbt!]
	\centering
		\includegraphics[width=0.7\textwidth]{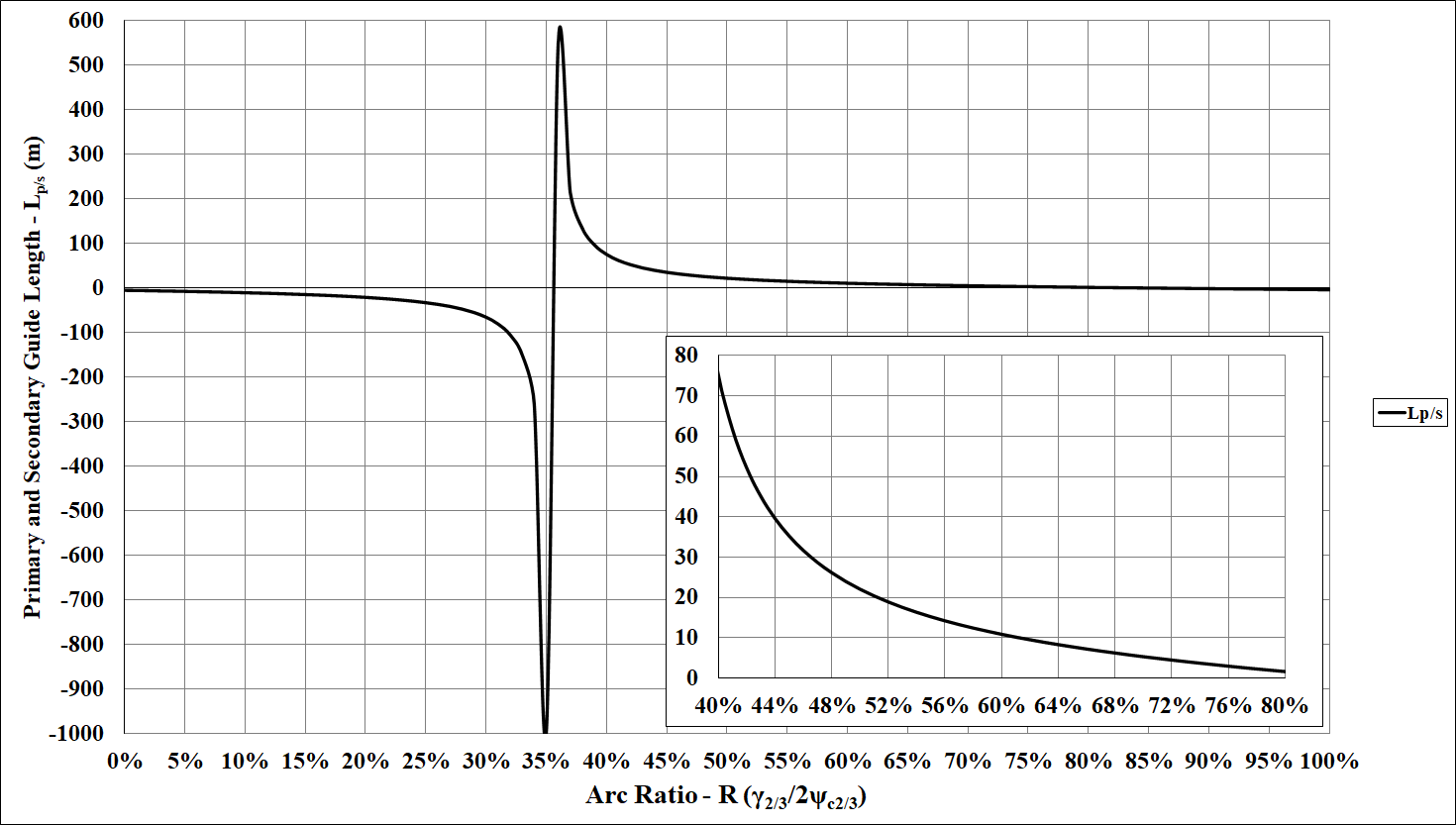}
	\caption{A plot of primary and secondary straight guide length versus arc ratio values, with values coming from Equation \ref{L_P_S} and correspond to cases with $\rho = 2000$~$m$. Equal curvatures corresponding to denominator singularities at an arc ratio of $35.35\%$. An additional graphic shows a close-up in the $40$~$\%-80$~$\%$ arc ratio range, which configurations have been used in the second set of simulations. }
\label{IV_L_ps}
\end{figure}

\begin{figure}[hbt!]
	\centering
		\includegraphics[width=0.7\textwidth]{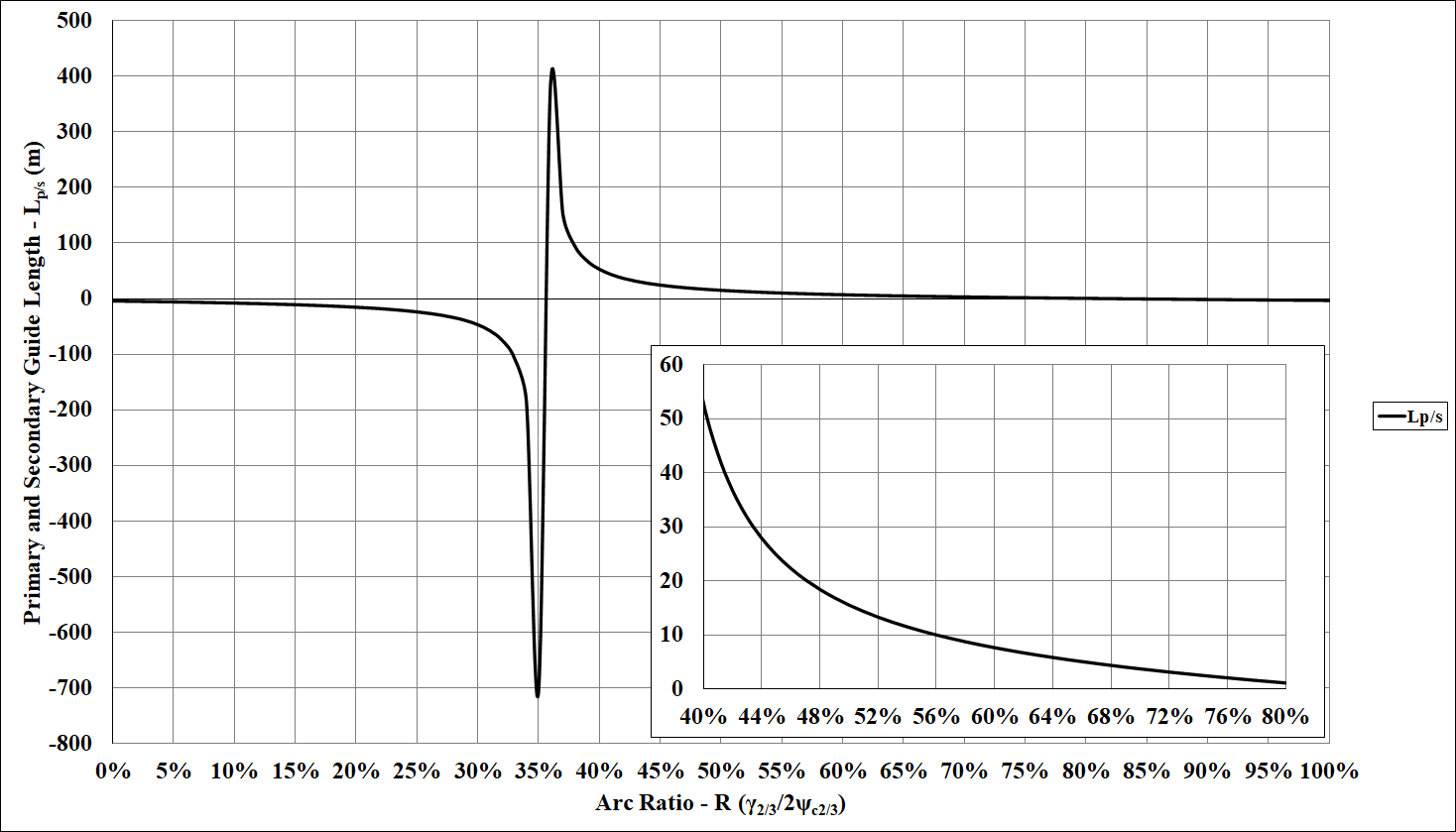}
	\caption{A plot of primary and secondary straight guide length versus arc ratio values, which values come from Equation \ref{L_P_S} and correspond to cases with $\rho = 1000$~$m$. Equal curvatures correspond to denominator singularities at an arc ratio of $35.35$~$\%$. An additional graphic shows a close-up in the $40$~$\%-80$~$\%$ arc ratio range, which configurations have been used in the second set of simulations. }
\label{V_L_ps}
\end{figure}

These scenarios are graphically presented in Figures \ref{IV_L_ps} and \ref{V_L_ps}, where, as mentioned before, a singularity in the primary and secondary straight guides is found at an arc ratio of $35.35$~$\%$ in an equal curvature S-shaped guide system. We took a close-up of the range between $40$~$\%$ and $80$~$\%$ on both figures to stress the simulation cases present in Table \ref{table_L_P_L_S}.

The last sequence of simulations comes from a deeper analysis of cases \textbf{II$^c$} and \textbf{IV$^c$}, in which the main characteristics are in Table \ref{tbl1}.  As previously said, simulation cases of Tables \ref{tbl1} and \ref{table_L_P_L_S} are performed by taking all coating indexes as $m=3$ and with a thermal ILL profile as the source. Here, we intend to change these indexes slightly and gradually from end system guides to the primary straight guide, according to the principle described in Equation \ref{escala_m}. We chose three different scenarios with combinations of two values of coating index, i.e., with $m=3.0$ and $m=2.5$, $m=3.0$ and $m=2.0$, and $m=2.5$ and $m=2.0$. Cases of \textbf{II$^c$} are in Table \ref{M_DIF_CASES_II} and of \textbf{IV$^c$} in Table \ref{M_DIF_CASES_IV}. All cases are carried out by those three types of ILL sources, as previously described.

Both tables contain values for each curved guide's characteristic wavelengths and parameters $\lambda^{\prime}$ for each curved guide. Depending on coating index values, parameters $\lambda^{\prime}$ might diverge, otherwise, the AD formalism provides the same guide neutron transport for cases \textbf{A} and \textbf{B} for a specter of a wavelength range between $\lambda_c$ and $\lambda^{\prime}$. 

Concerning this aspect, it was observed that among seven coating indexes configurations, case \textbf{D} is the most appropriate scenario to keep neutron transport efficiency using lower super mirror indexes along with an S-shaped guide system. In the next section, there is a comparison of these configurations next to their fluxes at the end of the guide system. In this set of simulations, we also tested wavelength cutoff properties according to Equations \ref{new_lambda_cut}, which allows comparing cutoffs of S-shaped guides with different super mirror indexes.

\begin{table}[hbt!]
\caption{Configuration of the third and last sequence of simulations. They correspond to seven S-shaped guide system coating arrangements, given from \textbf{A} to \textbf{G}. These configurations are then subdivided into the other three series of simulations according to super mirror index values (subcases \textbf{i}, \textbf{ii}, and \textbf{iii}). Columns $m_1$, $m_{out2}$, $m_{out3}$, $m_{in2}$, $m_{in3}$, and $m_{4}$ correspond to index values of primary straight guide all surfaces, first curved guide concave surface, second curved guide concave surface, first curved guide convex surface, second curved guide convex surface and secondary straight guide all surfaces, respectively. Columns of $\lambda_{c1}$, $\lambda_{c1}^{\prime}$, $\lambda_{c2}$ and $\lambda_{c2}^{\prime}$ show values of characteristic wavelength and parameter $\lambda^{\prime}$ values for the first and second curved guide, respectively. These values correspond to configuration \textbf{II} parameters (see Table \ref{tbl1}).}\label{M_DIF_CASES_II}
\begin{tabular*}{\tblwidth}{@{} LLLLLLLLLLL@{} }
\toprule
{\scriptsize\textbf{Case}}      & $m_{1}$ & $m_{out2}$ & $m_{out3}$ & $m_{in2}$ & $m_{in3}$ & $m_{4}$ & $\lambda_{c1}${\scriptsize(\AA)} & $\lambda_{c1}^{\prime}${\scriptsize(\AA)} & $\lambda_{c2}${\scriptsize(\AA)} & $\lambda_{c2}^{\prime}${\scriptsize(\AA)} \\
\midrule 
\textbf{A$^{i}$}   & 3.0       & 3.0          & 3.0          & 3.0         & 3.0         & 3.0      & 0.96           & $\infty$                & 1.36           & $\infty$                \\
\textbf{B$^{i}$}   & 3.0       & 3.0          & 3.0          & 3.0         & 3.0         & 2.5    & 0.96           & $\infty$                & 1.36           & $\infty$                \\
\textbf{C$^{i}$}   & 3.0       & 3.0          & 3.0          & 3.0         & 2.5       & 2.5    & 0.96           & $\infty$                & 1.36           & 2.46                    \\
\textbf{D$^{i}$}   & 3.0       & 3.0          & 3.0          & 2.5       & 2.5       & 2.5    & 0.96           & 1.74                    & 1.36           & 2.46                    \\
\textbf{E$^{i}$}   & 3.0       & 3.0          & 2.5        & 2.5       & 2.5       & 2.5    & 0.96           & 1.74                    & 1.63           & $\infty$                \\
\textbf{F$^{i}$}   & 3.0       & 2.5        & 2.5        & 2.5       & 2.5       & 2.5    & 1.16           & $\infty$                & 1.63           & $\infty$                \\
\textbf{G$^{i}$}   & 2.5     & 2.5        & 2.5        & 2.5       & 2.5       & 2.5    & 1.16           & $\infty$                & 1.63           & $\infty$                \\
\textbf{A$^{ii}$}  & 3.0       & 3.0          & 3.0          & 3.0         & 3.0         & 3.0      & 0.96           & $\infty$                & 1.36           & $\infty$                \\
\textbf{B$^{ii}$}  & 3.0       & 3.0          & 3.0          & 3.0         & 3.0         & 2.0     & 0.96           & $\infty$                & 1.36           & $\infty$                \\
\textbf{C$^{ii}$}  & 3.0       & 3.0          & 3.0          & 3.0         & 2.0        & 2.0     & 0.96           & $\infty$                & 1.36           & 1.83                    \\
\textbf{D$^{ii}$}  & 3.0       & 3.0          & 3.0          & 2.0        & 2.0        & 2.0     & 0.96           & 1.29                    & 1.36           & 1.83                    \\
\textbf{E$^{ii}$}  & 3.0       & 3.0          & 2.0         & 2.0        & 2.0        & 2.0     & 0.96           & 1.29                    & 2.04           & $\infty$                \\
\textbf{F$^{ii}$}  & 3.0       & 2.0         & 2.0         & 2.0        & 2.0        & 2.0     & 1.45           & $\infty$                & 2.04           & $\infty$                \\
\textbf{G$^{ii}$}  & 2.0      & 2.0         & 2.0         & 2.0        & 2.0        & 2.0     & 1.45           & $\infty$                & 2.04           & $\infty$                \\
\textbf{A$^{iii}$} & 2.5     & 2.5        & 2.5        & 2.5       & 2.5       & 2.5    & 1.16           & $\infty$                & 1.63           & $\infty$                \\
\textbf{B$^{iii}$} & 2.5     & 2.5        & 2.5        & 2.5       & 2.5       & 2.0     & 1.16           & $\infty$                & 1.63           & $\infty$                \\
\textbf{C$^{iii}$} & 2.5     & 2.5        & 2.5        & 2.5       & 2.0        & 2.0     & 1.16           & $\infty$                & 1.63           & 2.72                   \\
\textbf{D$^{iii}$} & 2.5     & 2.5        & 2.5        & 2.0        & 2.0        & 2.0     & 1.16           & 1.93                    & 1.63           & 2.72                    \\
\textbf{E$^{iii}$} & 2.5     & 2.5        & 2.0         & 2.0        & 2.0        & 2.0     & 1.16           & 1.93                    & 2.04           & $\infty$                \\
\textbf{F$^{iii}$} & 2.5     & 2.0         & 2.0         & 2.0        & 2.0        & 2.0     & 1.45           & $\infty$                & 2.04           & $\infty$                \\
\textbf{G$^{iii}$} & 2.0      & 2.0         & 2.0         & 2.0        & 2.0        & 2.0     & 1.45           & $\infty$                & 2.04           & $\infty$  \\
\bottomrule
\end{tabular*}
\vspace{-0.4cm}
\begin{flushleft}
{\footnotesize{$^i$ Combination of super mirror indexes of $m=3.0$ and $m=2.5$}}\\
{\footnotesize{$^{ii}$ Combination of super mirror indexes of $m=3.0$ and $m=2.0$}}\\
{\footnotesize{$^{iii}$ Combination of super mirror indexes of $m=2.5$ and $m=2.0$}}
\end{flushleft}
\end{table}

\begin{table}[hbt!]
\caption{Configuration of the third and last sequence of simulations. They correspond to seven S-shaped guide system coating arrangements, given from \textbf{A} to \textbf{G}. These configurations are then subdivided into the other three series of simulations according to super mirror index values (subcases \textbf{i}, \textbf{ii}, and \textbf{iii}). Columns $m_1$, $m_{out2}$, $m_{out3}$, $m_{in2}$, $m_{in3}$, and $m_{4}$ correspond to index values of primary straight guide all surfaces, first curved guide concave surface, second curved guide concave surface, first curved guide convex surface, second curved guide convex surface and secondary straight guide all surfaces, respectively. Columns of $\lambda_{c1}$, $\lambda_{c1}^{\prime}$, $\lambda_{c2}$ and $\lambda_{c2}^{\prime}$ show values of characteristic wavelength and parameter $\lambda^{\prime}$ values for the first and second curved guide, respectively. These values correspond to configuration \textbf{IV} parameters (see Table \ref{tbl1}).}\label{M_DIF_CASES_IV}
\begin{tabular*}{\tblwidth}{@{} LLLLLLLLLLL@{} }
\toprule
{\scriptsize\textbf{Case}}      & $m_{1}$ & $m_{out2}$ & $m_{out3}$ & $m_{in2}$ & $m_{in3}$ & $m_{4}$ & $\lambda_{c1}${\scriptsize(\AA)} & $\lambda_{c1}^{\prime}${\scriptsize(\AA)} & $\lambda_{c2}${\scriptsize(\AA)} & $\lambda_{c2}^{\prime}${\scriptsize(\AA)} \\
\midrule 
\textbf{A$^{i}$}   & 3.0       & 3.0          & 3.0          & 3.0         & 3.0         & 3.0      & 1.36           & $\infty$                & 1.36           & $\infty$                \\
\textbf{B$^{i}$}   & 3.0       & 3.0          & 3.0          & 3.0         & 3.0         & 2.5    & 1.36           & $\infty$                & 1.36           & $\infty$                \\
\textbf{C$^{i}$}   & 3.0       & 3.0          & 3.0          & 3.0         & 2.5       & 2.5    & 1.36           & $\infty$                & 1.36           & 2.46                    \\
\textbf{D$^{i}$}   & 3.0       & 3.0          & 3.0          & 2.5       & 2.5       & 2.5    & 1.36           & 2.46                    & 1.36           & 2.46                    \\
\textbf{E$^{i}$}   & 3.0       & 3.0          & 2.5        & 2.5       & 2.5       & 2.5    & 1.36           & 2.46                    & 1.63           & $\infty$                \\
\textbf{F$^{i}$}   & 3.0       & 2.5        & 2.5        & 2.5       & 2.5       & 2.5    & 1.63           & $\infty$                & 1.63           & $\infty$                \\
\textbf{G$^{i}$}   & 2.5     & 2.5        & 2.5        & 2.5       & 2.5       & 2.5    & 1.63           & $\infty$                & 1.63           & $\infty$                \\
\textbf{A$^{ii}$}  & 3.0       & 3.0          & 3.0          & 3.0         & 3.0         & 3.0      & 1.36           & $\infty$                & 1.36           & $\infty$                \\
\textbf{B$^{ii}$}  & 3.0       & 3.0          & 3.0          & 3.0         & 3.0         & 2.0     & 1.36           & $\infty$                & 1.36           & $\infty$                \\
\textbf{C$^{ii}$}  & 3.0       & 3.0          & 3.0          & 3.0         & 2.0        & 2.0     & 1.36           & $\infty$                & 1.36           & 1.83                    \\
\textbf{D$^{ii}$}  & 3.0       & 3.0          & 3.0          & 2.0        & 2.0        & 2.0     & 1.36           & 1.83                    & 1.36           & 1.83                    \\
\textbf{E$^{ii}$}  & 3.0       & 3.0          & 2.0         & 2.0        & 2.0        & 2.0     & 1.36           & 1.83                    & 2.04           & $\infty$                \\
\textbf{F$^{ii}$}  & 3.0       & 2.0         & 2.0         & 2.0        & 2.0        & 2.0     & 2.04           & $\infty$                & 2.04           & $\infty$                \\
\textbf{G$^{ii}$}  & 2.0      & 2.0         & 2.0         & 2.0        & 2.0        & 2.0     & 2.04           & $\infty$                & 2.04           & $\infty$                \\
\textbf{A$^{iii}$} & 2.5     & 2.5        & 2.5        & 2.5       & 2.5       & 2.5    & 1.63           & $\infty$                & 1.63           & $\infty$                \\
\textbf{B$^{iii}$} & 2.5     & 2.5        & 2.5        & 2.5       & 2.5       & 2.0     & 1.63           & $\infty$                & 1.63           & $\infty$                \\
\textbf{C$^{iii}$} & 2.5     & 2.5        & 2.5        & 2.5       & 2.0        & 2.0     & 1.63           & $\infty$                & 1.63           & 2.72                   \\
\textbf{D$^{iii}$} & 2.5     & 2.5        & 2.5        & 2.0        & 2.0        & 2.0     & 1.63           & 2.72                    & 1.63           & 2.72                    \\
\textbf{E$^{iii}$} & 2.5     & 2.5        & 2.0         & 2.0        & 2.0        & 2.0     & 1.63           & 2.72                    & 2.04           & $\infty$                \\
\textbf{F$^{iii}$} & 2.5     & 2.0         & 2.0         & 2.0        & 2.0        & 2.0     & 2.04           & $\infty$                & 2.04           & $\infty$                \\
\textbf{G$^{iii}$} & 2.0      & 2.0         & 2.0         & 2.0        & 2.0        & 2.0     & 2.04           & $\infty$                & 2.04           & $\infty$  \\
\bottomrule
\end{tabular*}
\vspace{-0.4cm}
\begin{flushleft}
{\footnotesize{$^i$ Combination of super mirror indexes of $m=3.0$ and $m=2.5$}}\\
{\footnotesize{$^{ii}$ Combination of super mirror indexes of $m=3.0$ and $m=2.0$}}\\
{\footnotesize{$^{iii}$ Combination of super mirror indexes of $m=2.5$ and $m=2.0$}}
\end{flushleft}
\end{table}

\section{Results and discussions}
\label{RAD}
\begin{figure}[hbt!]
	\centering
		\includegraphics[width=0.7\textwidth]{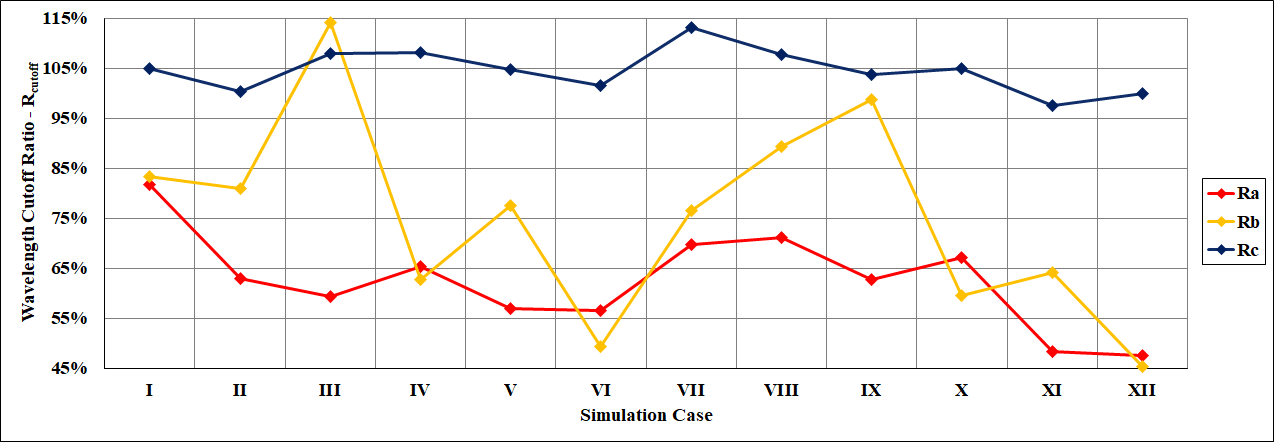}
	\caption{A plot of wavelength cutoff ratios versus the first sequence of simulation cases from \textbf{I} to \textbf{XII}. The ordinate axis represents ratios of simulated and theoretical wavelength cutoff, with values close to 100$~$\% indicating better fit between proposed simulations and literature. Blue, yellow and red curves stand for subcases \textbf{a.}, \textbf{b.}, and \textbf{c.}, respectively.}
\label{R_CUTOFF}
\end{figure}

The first set of simulations, shown in Table \ref{tbl1}, is carried out and the results are presented in Table \ref{VACA}. In complementarity, wavelength values are displayed graphically in Figure \ref{R_CUTOFF}.

In such a table, we present cases \textbf{I} to \textbf{XII} with triplet subcases that characterize both curved guide arcs. Since configurations, a and b are symmetrical, cases with equal curvature possess the same $Z$ displacement values. The second and third column contains curved guides arc ratios for both arched guides that compose the S-shaped guide. A comparison of theoretical and simulated wavelength cutoff denoted as $\lambda^{T}_{cut}$ and $\lambda^{S}_{cut}$, respectively, is summarized in the last column, through their ratio $R_\lambda$. 

Values of $R_\lambda$ are displayed in Figure \ref{R_CUTOFF}, where red, yellow and blue points and lines represent, respectively, subcases \textbf{a.}, \textbf{b.}, and \textbf{c.}. It is tacit that configurations \textbf{a.} and \textbf{b.}, despite their LoS exclusion, do not guarantee the wavelength cutoff properly. Only configuration \textbf{c.} allows arcs long enough to impose the cutoff.

\begin{table}[hbt!]
\caption{The result values and parameters of the first set of simulations. The second and third columns show that arc ratios of the first and second curved guides. The column $Z$ indicates vertical (or horizontal) displacement correspondent to the S-shaped guide configuration case. In the columns $\lambda_{cut}^{T}$ and $\lambda_{cut}^{S}$, theoretical and simulated values of the wavelength cutoff are displayed. Since theoretical values do not depend on arc values, there is only one value for all three cases \textbf{a.}, \textbf{b.}, and \textbf{c.}. The last column shows ratios of the previous two columns parameters, i.e., $\lambda_{cut}^{T}$ and $\lambda_{cut}^{S}$.}\label{VACA}
\begin{tabular*}{\tblwidth}{@{} LCCLLLL@{} }
\toprule
\textbf{Case}     & \textbf{$\gamma_2/2\Psi_{c2}$ $(\%)$} & \textbf{$\gamma_3/2\Psi_{c3}$ $(\%)$} & \textbf{$Z(m)$} & \textbf{$\lambda^{T}_{cut}$} (\AA) & \textbf{$\lambda^{S}_{cut}$} (\AA)& \textbf{$R_\lambda$ $(\%)$} \\ 
\midrule 
\textbf{I$^a$}    & 82.92                       & 30.36                        & 0.16            & \multirow{3}{*}{0.68}                 & 0.56                                  & 81.61                  \\ 
\textbf{I$^b$}    & 30.36                        & 82.92                        & 0.16            &                                       & 0.57                                  & 83.39                  \\
\textbf{I$^c$}    & 82.92                        & 82.92                        & 0.28            &                                       & 0.71                                  & 104.95                 \\ 
\textbf{II$^a$}   & 88.39                       & 21.80                        & 0.17            & \multirow{3}{*}{0.79}                 & 0.49                                  & 62.82                  \\ 
\textbf{II$^b$}   & 35.82                        & 75.46                        & 0.14            &                                       & 0.64                                  & 80.95                  \\ 
\textbf{II$^c$}   & 88.39                        & 75.46                        & 0.27            &                                       & 0.79                                  & 100.26                 \\ 
\textbf{III$^a$}  & 92.28                        & 12.36                        & 0.17            & \multirow{3}{*}{0.86}                 & 0.51                                  & 59.37                  \\ 
\textbf{III$^b$}  & 39.72                        & 67.61                        & 0.12            &                                       & 0.98                                  & 114.14                 \\ 
\textbf{III$^c$}  & 92.28                       & 67.61                        & 0.26            &                                       & 0.93                                  & 107.92                 \\ 
\textbf{IV$^a$}   & 81.94                        & 28.28                        & 0.15            & \multirow{3}{*}{0.96}                 & 0.63                                  & 65.30                  \\  
\textbf{IV$^b$}   & 28.28                        & 81.94                        & 0.15            &                                       & 0.60                                  & 62.74                  \\ 
\textbf{IV$^c$}   & 81.94                        & 81.94                        & 0.27            &                                       & 1.04                                  & 108.18                 \\ 
\textbf{V$^a$}    & 87.41                       & 18.87                        & 0.16            & \multirow{3}{*}{1.11}                 & 0.63                                  & 56.98                  \\ 
\textbf{V$^b$}    & 33.75                        & 74.12                        & 0.13            &                                       & 0.86                                  & 77.53                  \\ 
\textbf{V$^c$}    & 87.41                        & 74.12                        & 0.26            &                                       & 1.17                                  & 104.80                 \\ 
\textbf{VI$^a$}   & 80.60                        & 25.36                        & 0.14            & \multirow{3}{*}{1.36}                 & 0.77                                  & 56.62                  \\
\textbf{VI$^b$}   & 25.36                        & 80.60                        & 0.14            &                                       & 0.67                                  & 49.22                  \\  
\textbf{VI$^c$}   & 80.60                        & 80.60                        & 0.26            &                                       & 1.38                                  & 101.52                 \\ 
\textbf{VII$^a$}  & 78.41                        & 25.36                        & 0.14            & \multirow{3}{*}{0.68}                 & 0.47                                  & 69.63                  \\ 
\textbf{VII$^b$}  & 20.36                        & 80.60                        & 0.14            &                                       & 0.52                                  & 76.49                  \\ 
\textbf{VII$^c$}  & 78.41                        & 80.60                        & 0.25            &                                       & 0.77                                  & 113.07                 \\
\textbf{VIII$^a$} & 83.88                        & 14.73                        & 0.15            & \multirow{3}{*}{0.79}                 & 0.56                                  & 71.16                  \\ 
\textbf{VIII$^b$} & 25.82                        & 72.29                        & 0.12            &                                       & 0.70                                  & 89.22                  \\ 
\textbf{VIII$^c$} & 83.88                        & 72.29                        & 0.25            &                                       & 0.85                                  & 107.75                 \\ 
\textbf{IX$^a$}   & 87.78                        & 2.36                         & 0.15            & \multirow{3}{*}{0.86}                 & 0.54                                  & 62.72                  \\
\textbf{IX$^b$}   & 29.72                        & 63.35                        & 0.10            &                                       & 0.85                                  & 98.74                  \\ 
\textbf{IX$^c$}   & 87.78                        & 63.35                       & 0.23            &                                       & 0.89                                  & 103.80                 \\ 
\textbf{X$^a$}    & 75.86                        & 21.21                        & 0.12            & \multirow{3}{*}{0.96}                 & 0.65                                  & 67.20                  \\ 
\textbf{X$^b$}    & 14.14                        & 78.78                        & 0.13            &                                       & 0.57                                  & 59.60                  \\ 
\textbf{X$^c$}    & 75.86                        & 78.78                        & 0.24            &                                       & 1.01                                  & 104.89                 \\ 
\textbf{XI$^a$}   & 81.33                        & 8.87                         & 0.13            & \multirow{3}{*}{1.11}                 & 0.54                                  & 48.37                  \\ 
\textbf{XI$^b$}   & 19.61                        & 69.86                        & 0.11            &                                       & 0.71                                  & 64.06                  \\ 
\textbf{XI$^c$}   & 81.33                        & 69.86                        & 0.23            &                                       & 1.09                                  & 97.58                  \\ 
\textbf{XII$^a$}  & 72.56                        & 15.36                        & 0.11            & \multirow{3}{*}{1.36}                 & 0.65                                  & 47.46                  \\ 
\textbf{XII$^b$}  & 5.36                         & 76.35                        & 0.12            &                                       & 0.62                                  & 45.29                  \\ 
\textbf{XII$^c$}  & 72.56                       & 76.35                        & 0.22            &                                       & 1.36                                  & 99.86                  \\ 
\bottomrule
\end{tabular*}
\vspace{-0.4cm}
\begin{flushleft}
{\footnotesize{$^a$ $\gamma_2=\varphi+\varphi_{p}^{\prime\prime}$,   $\gamma_3=\varphi-\varphi_{s}^{\prime}$}}\\

{\footnotesize{$^b$ $\gamma_2=\varphi-\varphi_{p}^{\prime}$,   $\gamma_3=\varphi+\varphi_{s}^{\prime\prime}$}}\\

{\footnotesize{$^c$ $\gamma_2=\varphi+\varphi_{p}^{\prime\prime}$,   $\gamma_3=\varphi+\varphi_{s}^{\prime\prime}$}}
\end{flushleft}
\end{table}

According to these results, we have proposed a sequel of simulations to determine which arc ratio imposes wavelength cutoff to transported neutron through an S-shaped guide system. From Equation \ref{L_P_S}, we can get different arc ratios through variation of primary and secondary straight guide lengths or vice versa. A set of simulation cases is in Table \ref{table_L_P_L_S}, where straight guide lengths are obtained for twenty-one arc ratios from $40$~$\%$ to $80$~$\%$ in steps of $2$~$\%$.

\begin{figure}[hbt!]
	\centering
		\includegraphics[width=0.7\textwidth]{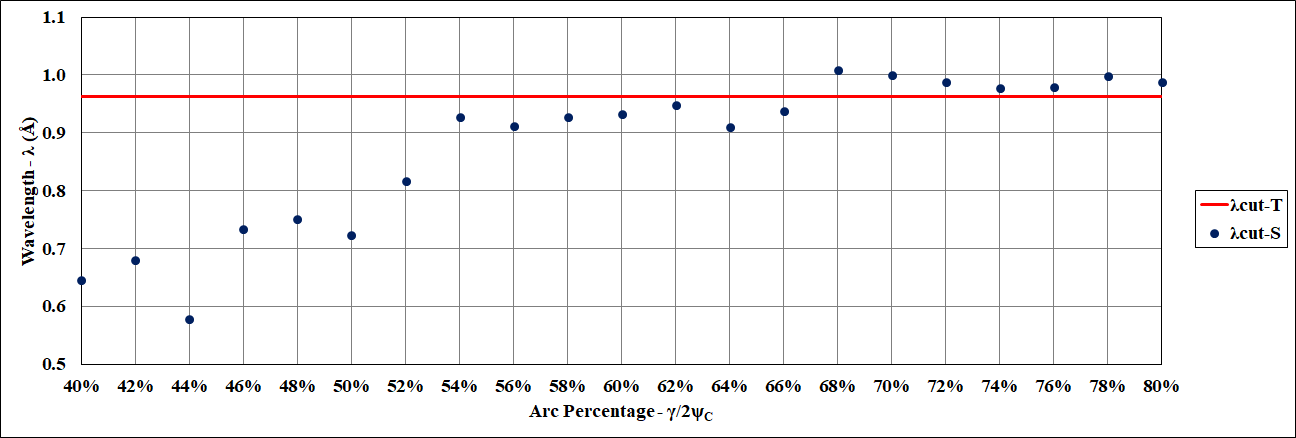}
\caption{A plot of wavelength cutoff values versus arc ratios. The ordinate axis represents wavelength values and the abscissa axis stands for ratios raging from $40$~$\%$ to $80$~$\%$. The results correspond to cases shown in Table \ref{table_L_P_L_S}, where each simulation configuration is gradually modified by an addition of $2\%$ in the ratio value. The red curve stands for a theoretical constant cutoff value, and the blue dots are simulated cutoff data. These results correspond to simulations of cases with $\rho = 2000$~$m$.}
\label{IV_lambda_cut}
\end{figure}

\begin{figure}[hbt!]
	\centering
		\includegraphics[width=0.5\textwidth]{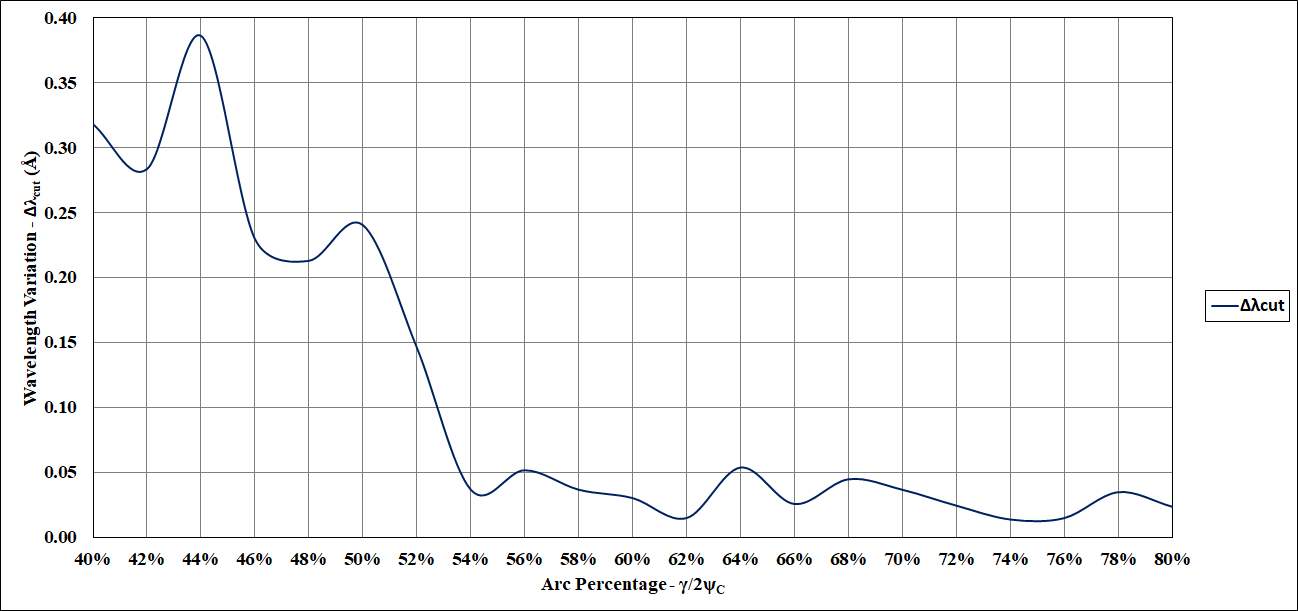}
\caption{A plot of the modular difference between theoretical and simulated wavelength cutoff values versus arc ratios. The ordinate axis represents wavelength values and the abscissa axis stands for ratios from $40$~$\%$ to $80$~$\%$ with steps of $2$~$\%$. Results correspond to cases with $\rho = 2000$~$m$, according to Table \ref{table_L_P_L_S}. The blue curve stands for the modular difference between two values of wavelength cutoff, so that the tendency to decay represents an improved adjustment of the theory and the presented simulations.}
\label{IV_dentalambda_cut}
\end{figure}

\begin{figure}[hbt!]
	\centering
		\includegraphics[width=0.7\textwidth]{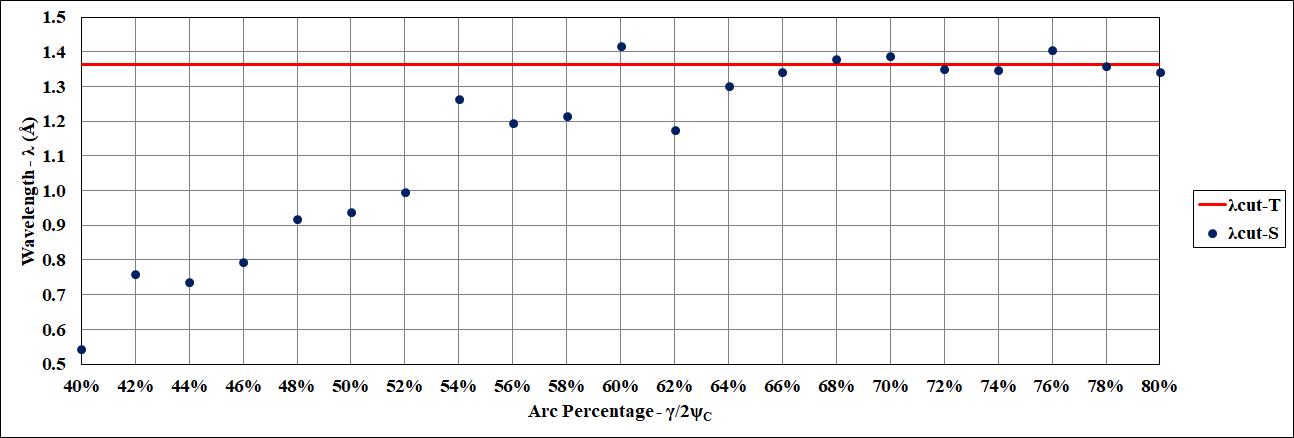}
\caption{A plot of wavelength cutoff values versus arc ratios. The ordinate represents wavelength and the abscissa stands for ratios raging from $40$~$\%$ to $80$~$\%$. The results correspond to cases shown in Table \ref{table_L_P_L_S}, where each simulation configuration is gradually modified by an addition of $2$~$\%$ in the ratio value. The red curve stands for a theoretical constant cutoff value, and the blue dots are simulated cutoff data. These results correspond to simulations of cases with $\rho = 1000$~$m$.}
\label{V_lambda_cut}
\end{figure}

\begin{figure}[hbt!]
	\centering
		\includegraphics[width=0.7\textwidth]{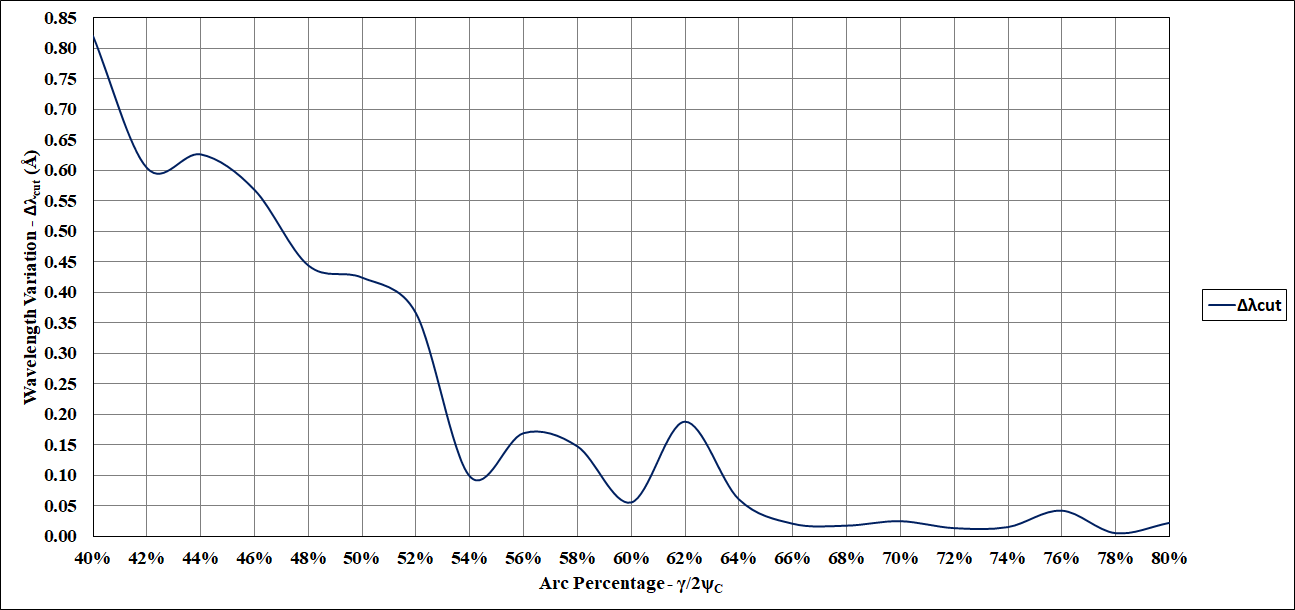}
\caption{A plot of the modular difference between theoretical and simulated wavelength cutoff values versus arc ratios. The ordinate axis represents wavelength values and the abscissa axis stands for ratios raging from $40$~$\%$ to $80$~$\%$ with steps of $2$~$\%$. The results correspond to the cases with $\rho = 1000$~$m$, according to Table \ref{table_L_P_L_S}. The blue curve stands for the modular difference between two values of wavelength cutoff, so that the tendency to decay represents an improved adjustment of the theory and the presented simulations. }
\label{V_deltalambda_cut}
\end{figure}

Simulations of Table \ref{table_L_P_L_S} are carried out and results are presented in two graphic pairs, each one for a different curved guide curvature, i.e., $\rho = 2000$~$m$ and $\rho = 1000$~$m$, since $\rho_2=\rho_3$. Here, Figures \ref{IV_lambda_cut} and \ref{IV_dentalambda_cut} represent simulations with $\rho = 2000$~$m$ and Figures \ref{V_lambda_cut} and \ref{V_deltalambda_cut} the other with $\rho = 1000$~$m$.

In Figures \ref{IV_lambda_cut} and \ref{V_lambda_cut}, there are horizontal red lines that stand for theoretical wavelength cutoff values of each configuration. Blue dots indicate correspondent cutoff values from simulated cases with different arc ratios. On the other hand, it is observed in Figures \ref{IV_dentalambda_cut} and \ref{V_deltalambda_cut} the modulus of both theoretical and simulated cutoff difference values, i.e., $\Delta\lambda_{cut}=|\lambda^{T}_{cut}-\lambda^{S}_{cut}|$.

\begin{figure}[hbt!]
	\centering
		\includegraphics[width=0.7\textwidth]{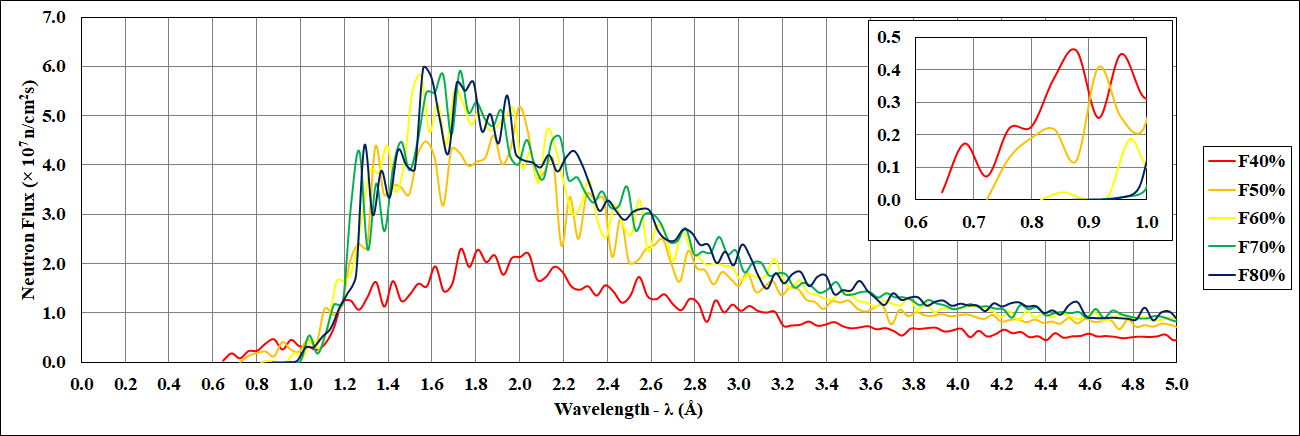}
\caption{The plot shows respective flux profiles at the end of the S-shaped guide system according to their arc rations. The ordinate and abscissa axes stand for neutron flux in $10^7 n/cm^2s$ units and wavelength in Angstroms, respectively. Red, orange, yellow, green, and blue curves represent arc ratio values of $40$~$\%$, $50$~$\%$, $60$~$\%$, $70$~$\%$, and $80$~$\%$, respectively. The close-up graph indicates that arc ratios $70$~$\%$ and $80$~$\%$ possess their wavelength cutoff values close to $0.96$ \AA, which is the theoretical cutoff according to Figure \ref{IV_lambda_cut}.}
\label{IV_40_50_60_70_80}
\end{figure}

\begin{figure}[hbt!]
	\centering
		\includegraphics[width=0.7\textwidth]{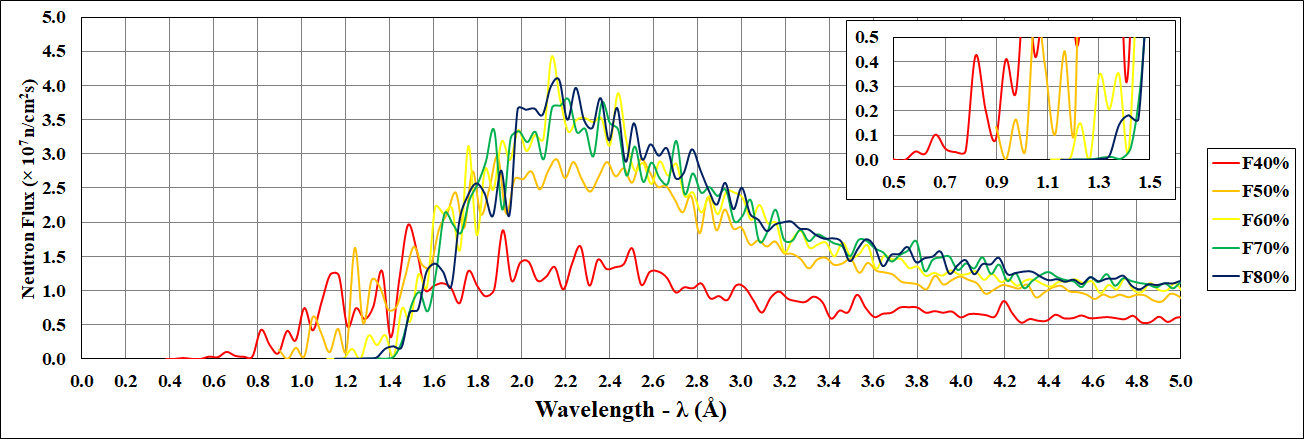}
\caption{The plot shows respective flux profiles at the end of the S-shaped guide system according to their arc rations. The ordinate and abscissa axes stand for neutron flux in $10^7 n/cm^2s$ units and wavelength in Angstroms, respectively. Red, orange, yellow, green, and blue curves represent arc ratio values of $40$~$\%$, $50$~$\%$, $60$~$\%$, $70$~$\%$, and $80$~$\%$, respectively. The close-up graph indicates that arc ratios $70$~$\%$ and $80$~$\%$ possess their wavelength cutoff values close to $1.36$ \AA, which is the theoretical cutoff according to Figure \ref{V_lambda_cut}.}
\label{V_40_50_60_70_80}
\end{figure}

Based on these results, it is possible to observe that as much as the arc ratio increases, the simulated wavelength cutoff approaches the theoretical one.  In other words, the $\Delta\lambda_{cut}$ tends to zero if $R_\lambda$ goes to one. In this scenario, the relation $\Delta\lambda_{cut}<0.05$ \AA$ $ is guaranteed for $R_\lambda>65$~$\%$. It is worth noting that, according to Figures \ref{IV_L_ps} and \ref{V_L_ps}, higher arc ratio values correspond to shorter primary and secondary straight guides. Therefore, S-shaped guide systems with wavelength cutoff are desirable from a neutron transportation point of view. Figures \ref{V_40_50_60_70_80} and \ref{IV_40_50_60_70_80} show neutron flux profiles at the end of S-shaped guide systems of this set of simulations, but only for cases of arc ratios of $40$~$\%$, $50$~$\%$, $60$~$\%$, $70$~$\%$ and $80$~$\%$. 

Table \ref{table_L_P_L_S_FLUX} shows flux values of these profiles. Here, one can see that longer arcs correspond to longer curved guides but to shorter straight guides. However, these cases provide higher flux values since the exclusion of LoS in S-shaped guides forces neutrons to go through longer paths. According to Equation \ref{Z} and Table \ref{table_L_P_L_S_FLUX}, we also note that higher flux cases provide longer vertical (or horizontal) displacement.

\begin{table}[hbt!]
\caption{Flux data results of the second run of simulations. Presented fluxes correspond only to tens of arc ratio whole sequence, i.e., $40$~$\%$, $50$~$\%$, $60$~$\%$, $70$~$\%$ and $80$~$\%$. Fluxes are given on a $\times 10^9 n/cm^2s$ basis and the second and third columns represent, respectively, simulations with $\rho= 2000$~$m$ and $\rho = 1000$~$m$. After flux values and between parenthesis, there are relative percentages of correspondent values next to the highest flux of the simulation series. That is, fluxes of arc ratios of $80$~$\%$ are the top flux values and are taken as $100$~$\%$.}\label{table_L_P_L_S_FLUX}
\begin{tabular*}{\tblwidth}{@{} LCC@{} }
\toprule
\multirow{2}{*}{$R_{2/3}$ $(\%)$} & \multicolumn{2}{c}{Flux $(\times 10^9 n/cm^2s)$} \\
                           & $\rho = 2000$~$m$   & $\rho = 1000$~$m$ $^{\dagger}$ \\
\midrule 

$40$ & $1.588$~$(46.11\%) $ & $ 1.544$~$(50.06\%)$ \\
$50$ & $2.860$~$(83.04\%) $ & $ 2.618$~$(84.89\%)$  \\
$60$ & $3.266$~$(94.83\%) $ & $ 2.954$~$(95.78\%)$  \\
$70$ & $3.431$~$(99.62\%) $ & $ 3.027$~$(98.15\%)$  \\
$80$ & $3.444$~$(100.00\%)$ & $ 3.084$~$(100.00\%)$  \\

\bottomrule
\end{tabular*}
\vspace{-0.2cm}
\begin{flushleft}
\footnotesize{$^{\dagger}$ $\rho = \rho_2 = \rho_3$}
\end{flushleft}

\end{table}

\begin{figure}[hbt!]
	\centering
		\includegraphics[width=0.7\textwidth]{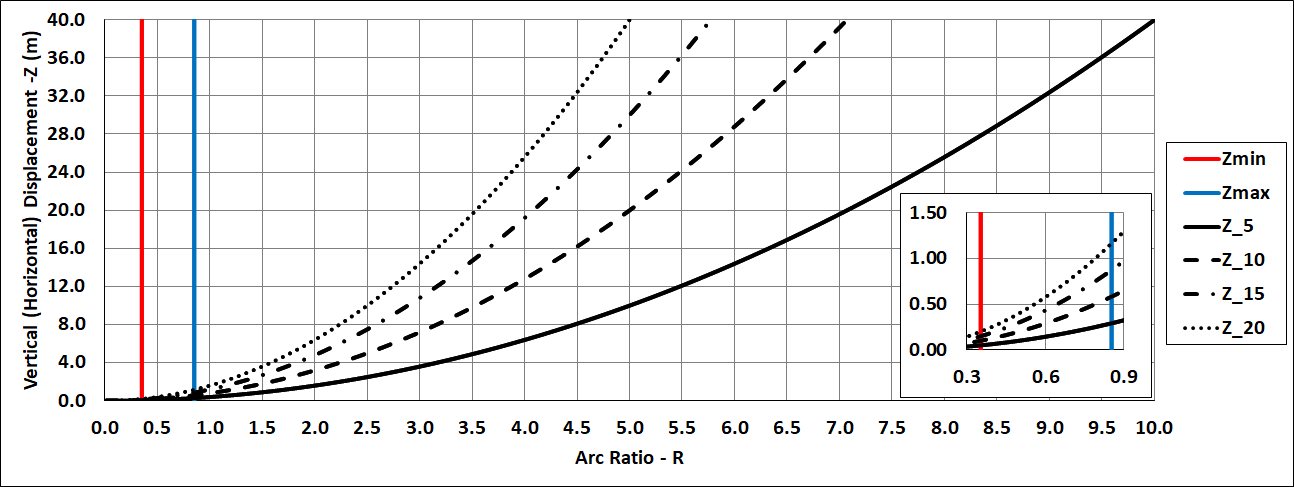}
	\caption{The plot of vertical (or horizontal) displacement versus arc ratio. The ordinate axis represents the parameter $Z$ and the abscissa axis the arc ratio. Vertical blue and red curves indicate maximum and minimum ratio values corresponding to the minimal S-shaped guide formalism of the present work. Solid, dashed, dotted-dashed, and dotted black curves stand for vertical (or horizontal) displacements of guide systems with widths of $5$, $10$, $15$, and $20$~$cm$, respectively. }
\label{Zmin_Zmax}
\end{figure}

There is in Figure \ref{Zmin_Zmax} a proper interval of arc ratios that guarantees LoS exclusion, between the vertical red and blue curves, corresponding to $35.36$~$\%$ and $85.36$~$\%$, respectively. The S-shaped guide displacement curve follows the equation $Z=8WR^2$ since guide curvatures are way longer than their correspondent lengths, i.e., for ($L_{2/3}~<<~\rho_{2/3}$). In this sense, both scenarios with equal curvature $\rho_2$ and $\rho_3$ possess the same plot to represent minimum and maximum $Z$ displacement values according to a minimal S-shaped guide system. Anyway, the present formalism may be violated if higher displacement values are needed. Otherwise, guide width can be changed to allow modifying the solid black curve parabola inclination of Figure \ref{Zmin_Zmax}, which corresponds to $Z$ displacement with $W = 5$~$cm$ and is a standard value for all accomplished simulation cases of this work. In addition, dashed, dotted-dashed, and dotted curves stand for a vertical (or horizontal) displacement of S-shaped curves with widths of $10$, $15$, and $20$~$cm$, respectively.

In these minimal regimes of LoS exclusion, a flux difference of about 2 times between arc ratios of $40$~$\%$ and $80$~$\%$ was observed. In addition, it is not possible to obtain higher displacements of about $29$~$cm$ as previously mentioned, the only way to obtain larger $Z$ values is by employing wider guides (larger $W$). Otherwise, if minimal LoS exclusion is not required, then longer displacements are available.  

The last set of simulations corresponds to  the cases presented in Tables \ref{M_DIF_CASES_II} and \ref{M_DIF_CASES_IV}, which correspond to unfolding the cases \textbf{II} and \textbf{IV}. Here, we reemphasize two aspects of these simulations. Firstly, we verified simulations from a wavelength cutoff point of view, i.e., the proposed cutoff expression for different surface indexes m WAS tested since we convert engineering aspects of neutron guides into geometrical ones (see Equation \ref{new_varphi}). Secondly, we verified fluxes at the end of simulated guide systems. From these results, we intend to identify super mirror coating arrangements that preserve fluxes close to ideal cases, i.e., all indexes equal and with maximum value. 

\begin{table}[hbt!]
\caption{Wavelength cutoff data from the last sequence of simulations. These results come from running Table \ref{M_DIF_CASES_II} cases for three different ILL sources, i.e., with cold, thermal, and hot spectra. The second column contains theoretical values of cutoff, while the posterior columns present simulated wavelength cutoff values for three runs of simulations with each source. Together with all simulated data, there is a percentage between parenthesis that stands for theoretical and simulated ratio percentage. This table of results corresponds to simulation case \textbf{II}, which configurations are in Table \ref{tbl1}}\label{CTH_II}
\begin{tabular*}{\tblwidth}{@{} LLCCC@{} }
\toprule
\multirow{2}{*}{\textbf{Case}} & \multirow{2}{*}{$\lambda_{cut}^{T}$ (\AA)} & \textbf{Cold}                                                     & \textbf{Thermal}                                                    & \textbf{Hot}                                                        \\
                               &                                                           & $\lambda_{cut}^{S}$ (\AA)($R_{\lambda}$) & $\lambda_{cut}^{S}$(\AA)($R_{\lambda}$) & $\lambda_{cut}^{S}$(\AA)($R_{\lambda}$) \\

\midrule 
\textbf{A$^{i}$}               & 0.79                                                      & 0.95  (120.30\%)                          & 0.84 (106.36\%)                          & 0.75 (94.73\%)                           \\
\textbf{B$^{i}$}               & 0.79                                                      & 0.90 (115.04\%)                          & 0.80 (102.05\%)                          & 0.83 (105.09\%)                          \\
\textbf{C$^{i}$}               & 0.79                                                      & 0.90 (114.22\%)                          & 0.90 (114.00\%)                          & 0.74 (94.09\%)                           \\
\textbf{D$^{i}$}               & 0.79                                                      & 0.94 (119.77\%)                          & 0.82 (104.21\%)                          & 0.85 (107.58\%)                          \\
\textbf{E$^{i}$}               & 0.83                                                      & 0.97 (117.02\%)                          & 0.85 (102.50\%)                          & 0.82 (99.04\%)                           \\
\textbf{F$^{i}$}               & 0.94                                                      & 1.00 (105.92\%)                          & 0.97 (102.63\%)                          & 0.91 (96.69\%)                           \\
\textbf{G$^{i}$}               & 0.94                                                      & 1.00 (106.03\%)                          & 0.97 (102.78\%)                          & 0.88 (92.94\%)                           \\
\textbf{A$^{ii}$}              & 0.79                                                      & 0.93 (117.76\%)                          & 0.82 (104.87\%)                          & 0.76 (97.18\%)                           \\
\textbf{B$^{ii}$}              & 0.79                                                      & 0.85 (108.47\%)                          & 0.88 (112.34\%)                          & 0.80 (101.07\%)                          \\
\textbf{C$^{ii}$}              & 0.79                                                      & 0.91 (115.34\%)                          & 0.83 (105.67\%)                          & 0.75 (95.09\%)                           \\
\textbf{D$^{ii}$}              & 0.79                                                      & 0.86 (109.27\%)                          & 0.84 (107.10\%)                          & 0.74 (94.63\%)                           \\
\textbf{E$^{ii}$}              & 0.87                                                      & 1.01 (116.10\%)                          & 0.88 (100.86\%)                          & 0.99 (113.87\%)                          \\
\textbf{F$^{ii}$}              & 1.18                                                      & 1.29 (109.17\%)                          & 1.11 (94.21\%)                           & 1.20 (102.00\%)                          \\
\textbf{G$^{ii}$}              & 1.18                                                      & 1.23 (104.28\%)                          & 1.17 (98.76\%)                           & 1.13 (95.39\%)                           \\
\textbf{A$^{iii}$}             & 0.94                                                      & 1.00 (106.28\%)                          & 0.95 (101.17\%)                          & 0.95 (100.85\%)                          \\
\textbf{B$^{iii}$}             & 0.94                                                      & 1.01 (107.49\%)                          & 0.95 (100.25\%)                          & 0.92 (96.99\%)                           \\
\textbf{C$^{iii}$}             & 0.94                                                      & 1.00 (105.59\%)                          & 0.93 (98.93\%)                           & 0.96 (102.06\%)                          \\
\textbf{D$^{iii}$}             & 0.94                                                      & 1.02 (107.55\%)                          & 0.95 (100.47\%)                          & 0.96 (102.08\%)                          \\
\textbf{E$^{iii}$}             & 1.01                                                      & 1.10 (109.49\%)                          & 1.02 (101.52\%)                          & 1.02 (101.46\%)                          \\
\textbf{F$^{iii}$}             & 1.18                                                      & 1.25 (106.11\%)                          & 1.17 (99.13\%)                           & 1.12 (94.97\%)                           \\
\textbf{G$^{iii}$}             & 1.18                                                      & 1.21 (102.45\%)                          & 1.11 (93.78\%)                           & 1.14 (96.78\%)  \\                        
 
\bottomrule
\end{tabular*}
\vspace{-0.4cm}
\begin{flushleft}
{\footnotesize{$^i$ Combination of super mirror indexes of $m=3.0$ and $m=2.5$}}\\
{\footnotesize{$^{ii}$ Combination of super mirror indexes of $m=3.0$ and $m=2.0$}}\\
{\footnotesize{$^{iii}$ Combination of super mirror indexes of $m=2.5$ and $m=2.0$}}
\end{flushleft}
\end{table}

Simulations are carried out and results, which concern wavelength cutoff, are displayed in Tables \ref{CTH_II} and \ref{CTH_IV}, where the former corresponds to \textbf{II} cases and the latter to \textbf{IV} cases. Both tables are composed of one column of theoretical wavelength cutoff, calculated according to Equation \ref{new_lambda_cut}, and three columns of simulation cutoff results. Here, each one of them relate to a different ILL source, namely cold, thermal and hot sources. To simplify data analysis, there are corresponding percentages of simulated cutoffs compared to theoretical ones. Such ratios stay after simulated cutoffs in between parentheses. An additional table is attached to evaluate the validity of Equation \ref{new_lambda_cut} employing compact statistical data. Such results are in Table \ref{STAT}. Results between simulated and theoretical wavelength cutoffs are counted in two ranges, namely, $\pm 5$~$\%$ and $\pm 10$~$\%$, i.e., if cutoff the ratios are between $ 95$~$\%$ and $ 105$~$\%$ or $ 90$~$\%$ and $ 110$~$\%$, respectively.

\begin{table}[hbt!]
\caption{Wavelength cutoff data from the last sequence of simulations. These results come from running Table \ref{M_DIF_CASES_IV} cases for three different ILL sources, i.e., with cold, thermal, and hot spectra. The second column contains theoretical values of cutoff, while the posterior columns present simulated wavelength cutoff values for three runs of simulations with each source. Together with all simulated data, there is a percentage between parenthesis that stands for theoretical and simulated ratio percentage. This table of results corresponds to simulation case \textbf{IV}, which configurations are in Table \ref{tbl1}}\label{CTH_IV}
\begin{tabular*}{\tblwidth}{@{} LLCCC@{} }
\toprule
\multirow{2}{*}{\textbf{Case}} & \multirow{2}{*}{$\lambda_{cut}^{T}$ (\AA)} & \textbf{Cold}                                                     & \textbf{Thermal}                                                    & \textbf{Hot}                                                        \\
                               &                                                           & $\lambda_{cut}^{S}$ (\AA)($R_{\lambda}$) & $\lambda_{cut}^{S}$(\AA)($R_{\lambda}$) & $\lambda_{cut}^{S}$(\AA)($R_{\lambda}$) \\

\midrule 
\textbf{A$^{i}$}               & 0.96                                                      & 1.02  (105.82\%)                          & 1.00 (103.59\%)                          & 0.96 (99.60\%)                           \\
\textbf{B$^{i}$}               & 0.96                                                      & 0.98 (101.93\%)                          & 0.98 (101.66\%)                          & 0.96 (99.60\%)                          \\
\textbf{C$^{i}$}               & 0.96                                                      & 0.98 (101.78\%)                          & 1.01 (104.46\%)                          & 0.95 (98.60\%)                           \\
\textbf{D$^{i}$}               & 0.96                                                      & 1.01 (105.21\%)                          & 0.96 (99.63\%)                          & 0.96 (99.76\%)                          \\
\textbf{E$^{i}$}               & 1.05                                                      & 1.12 (106.82\%)                          & 1.05 (100.63\%)                          & 1.03 (98.61\%)                           \\
\textbf{F$^{i}$}               & 1.16                                                      & 1.26 (108.79\%)                          & 1.16 (99.95\%)                          & 1.11 (95.93\%)                           \\
\textbf{G$^{i}$}               & 1.16                                                      & 1.19 (102.72\%)                          & 1.15 (99.37\%)                          & 1.10 (95.52\%)                           \\
\textbf{A$^{ii}$}              & 0.96                                                      & 1.01 (104.33\%)                          & 0.98 (102.24\%)                          & 0.99 (102.87\%)                           \\
\textbf{B$^{ii}$}              & 0.96                                                      & 1.04 (108.30\%)                          & 0.99 (102.93\%)                          & 1.00 (103.65\%)                          \\
\textbf{C$^{ii}$}              & 0.96                                                      & 1.05 (108.56\%)                          & 0.99 (103.12\%)                          & 0.97 (100.98\%)                           \\
\textbf{D$^{ii}$}              & 0.96                                                      & 1.02 (105.89\%)                          & 1.03 (106.57\%)                          & 0.98 (101.92\%)                           \\
\textbf{E$^{ii}$}              & 1.13                                                      & 1.23 (108.87\%)                          & 1.10 (97.26\%)                          & 1.05 (92.88\%)                          \\
\textbf{F$^{ii}$}              & 1.45                                                      & 1.49 (103.11\%)                          & 1.37 (95.04\%)                           & 1.41 (97.31\%)                          \\
\textbf{G$^{ii}$}              & 1.45                                                      & 1.47 (101.48\%)                          & 1.42 (98.34\%)                           & 1.36 (94.04\%)                           \\
\textbf{A$^{iii}$}             & 1.16                                                      & 1.22 (105.48\%)                          & 1.14 (98.21\%)                          & 1.19 (102.87\%)                          \\
\textbf{B$^{iii}$}             & 1.16                                                      & 1.25 (108.08\%)                          & 1.18 (101.70\%)                          & 1.16 (100.20\%)                           \\
\textbf{C$^{iii}$}             & 1.16                                                      & 1.16 (100.03\%)                          & 1.15 (99.15\%)                           & 1.12 (96.77\%)                          \\
\textbf{D$^{iii}$}             & 1.16                                                      & 1.20 (103.98\%)                          & 1.11 (96.20\%)                          & 1.13 (97.58\%)                          \\
\textbf{E$^{iii}$}             & 1.28                                                      & 1.32 (103.63\%)                          & 1.21 (94.74\%)                          & 1.28 (99.89\%)                          \\
\textbf{F$^{iii}$}             & 1.45                                                      & 1.52 (105.10\%)                          & 1.39 (96.53\%)                           & 1.41 (97.53\%)                           \\
\textbf{G$^{iii}$}             & 1.45                                                      & 1.49 (102.99\%)                          & 1.39 (96.51\%)                           & 1.35 (93.51\%)  \\                        

\bottomrule
\end{tabular*}
\vspace{-0.4cm}
\begin{flushleft}
{\footnotesize{$^i$ Combination of super mirror indexes of $m=3.0$ and $m=2.5$}}\\
{\footnotesize{$^{ii}$ Combination of super mirror indexes of $m=3.0$ and $m=2.0$}}\\
{\footnotesize{$^{iii}$ Combination of super mirror indexes of $m=2.5$ and $m=2.0$}}
\end{flushleft}
\end{table}

All simulations are carried out using $10^7$ rays in the MCSTAS software. In this scenario, the fluxes present an error of about $3$~$\%$. This percentage concerns the number of neutrons that reach a detector. In addition, our criteria for picking a wavelength cutoff are based on neutron counting of less than $1$~$\%$ from the profile peak.  Thus, we consider tests of $5$~$\%$ and $10$~$\%$ a proper method to verify wavelength cutoff equation validity. According to Table \ref{STAT}, we observe that cold source simulations possess less than $50$~$\%$  cutoff value, except simulations of Case IV. We believe that this discrepancy is due to a minor concentration of neutrons in the vicinity of the wavelength cutoff, which diminishes neutron counting and the statistics to determine neutrons of minimal wavelength. In other words, from Figure \ref{SOURCE} it is possible to notice that neutron counting near typical cutoffs, i.e., between $0.8$ and $1.5$ \AA, is much more prevalent for hot and thermal sources than for cold ones. Other scenarios show a good agreement between theory and simulations. 

\begin{table}[hbt!]
\caption{Compilation of wavelength cutoff statistical results of \textbf{II} and \textbf{IV} configuration simulations and carried out for three ILL sources. The presented numbers correspond individually for each case and source type, which is displayed in the first and second columns, respectively. The third column indicates that 21 simulations were run for different arrangement parameters. The last two columns show the number of simulated cases, in which wavelength cutoff values differ just $5$~$\%$ and $10$~$\%$ from the theoretical value, respectively. These numbers are followed by a percentage between parenthesis that indicates the percentage of cases of correspondent ranges.}\label{STAT}
\begin{tabular*}{\tblwidth}{@{} LLLLL@{} }
\toprule

\multirow{2}{*}{\textbf{Case}} & \multirow{2}{*}{\textbf{Source}} & \textbf{Number of}   & $|\Delta R_\lambda|^{*}<=5\%$ & $|\Delta R_\lambda|^{*}<=10\%$ \\
                               &                                  & \textbf{Simulations} & (Percentage)             & (Percentage)              \\

\midrule 
\multirow{3}{*}{\textbf{II}}   & \textbf{Cold}                    & 21                   & 2 (9.58\%)               & 13 (61.90\%)              \\
                               & \textbf{Thermal}                 & 21                   & 14 (66.67\%)             & 19 (90.48\%)              \\
                               & \textbf{Hot}                     & 21                   & 13 (61.90\%)             & 20 (95.24\%)              \\
\multirow{3}{*}{\textbf{IV}}   & \textbf{Cold}                    & 21                   & 10 (47.62\%)             & 21 (100.00\%)             \\
                               & \textbf{Thermal}                 & 21                   & 19 (90.48\%)             & 21 (100.00\%)             \\
                               & \textbf{Hot}                     & 21                   & 18 (85.71\%)             & 21 (100.00\%)            
     \\                        

\bottomrule
\end{tabular*}
\vspace{-0.4cm}
\begin{flushleft}
{\footnotesize{$^*$ $|\Delta R_\lambda| = |100$~$\%-R_\lambda|$}}

\end{flushleft}
\end{table}

The investigation of neutron transport efficiency is done through Tables \ref{FLUX_II} and \ref{FLUX_IV}, which correspond to cases \textbf{II} and \textbf{IV}, respectively. These tables are composed of three main columns that show neutron flux at the end of the S-shaped guide system and correspond to cold, thermal, and hot simulations. From the neutron intensity point of view, it is verified that colder neutron specters possess higher fluxes, since most of the hot and thermal neutrons are excluded from the profile by wavelength cutoff.

\begin{table}[hbt!]
\caption{Flux data results of the last run of simulations. Configurations of Table \ref{M_DIF_CASES_II} are carried out for three different ILL virtual sources, i.e., with cold, thermal, and hot neutron spectra. Cold and thermal fluxes, described respectively by the second and third columns, are given in $\times 10^9$~$n/cm^2s$, while hot results in the last column are in $\times 10^8$~$n/cm^2s$. Flux values come together with percentages between parenthesis that correspond to the ratio of each flux value next to the case A of each sequence. These percentages are just relative to cases \textbf{A} to \textbf{G} and their respective subcases, i.e., \textbf{i}, \textbf{ii}, and \textbf{iii}, and cases \textbf{A} are the standard comparison value (always with percentage $100$~$\%$). Results correspond to simulations of case \textbf{II} from Table \ref{tbl1}.}\label{FLUX_II}
\begin{tabular*}{\tblwidth}{@{} LLLL@{} }
\toprule
\multirow{3}{*}{\textbf{Case}} & \textbf{Cold}               & \textbf{Thermal}          & \textbf{Hot}              \\
                               & $F_{cur}(R_{flux}^{abs})$   & $F_{cur}(R_{flux}^{abs})$ & $F_{cur}(R_{flux}^{abs})$ \\
                               & $ (\times 10^{9}   n/cm^2s)$ & $ (\times 10^{9} n/cm^2s)$ & $ (\times 10^{8} n/cm^2s)$ \\

\midrule 
\textbf{A$^{i}$}               & 8.98 (100.00\%)             & 3.40 (100.00\%)           & 2.17 (100.00\%)           \\
\textbf{B$^{i}$}               & 8.99 (100.04\%)             & 3.42 (100.50\%)           & 2.14 (98.35\%)            \\
\textbf{C$^{i}$}               & 9.00 (100.22\%)             & 3.44 (101.13\%)           & 2.08 (95.73\%)            \\
\textbf{D$^{i}$}               & 9.00 (100.20\%)             & 3.37 (99.23\%)            & 2.17 (99.77\%)            \\
\textbf{E$^{i}$}               & 8.79 (97.84\%)              & 3.11 (91.45\%)            & 1.68 (77.53\%)            \\
\textbf{F$^{i}$}               & 8.72 (97.05\%)              & 2.94 (86.46\%)            & 1.50 (69.21\%)            \\
\textbf{G$^{i}$}               & 8.72 (97.05\%)              & 2.95 (86.77\%)            & 1.52 (69.78\%)            \\
\textbf{A$^{ii}$}              & 8.97 (100.00\%)             & 3.39 (100.00\%)           & 2.09 (100.00\%)           \\
\textbf{B$^{ii}$}              & 8.98 (100.08\%)             & 3.41 (100.44\%)           & 2.20 (105.42\%)           \\
\textbf{C$^{ii}$}              & 9.00 (100.30\%)             & 3.36 (99.08\%)            & 2.13 (101.99\%)           \\
\textbf{D$^{ii}$}              & 8.87 (98.82\%)              & 3.30 (97.30\%)            & 2.07 (99.44\%)            \\
\textbf{E$^{ii}$}              & 8.24 (91.86\%)              & 2.53 (74.57\%)            & 1.17 (56.33\%)            \\
\textbf{F$^{ii}$}              & 8.04 (89.62\%)              & 2.35 (69.22\%)            & 0.84 (40.13\%)            \\
\textbf{G$^{ii}$}              & 8.11 (90.45\%)              & 2.31 (69.15\%)            & 0.85 (40.69\%)            \\
\textbf{A$^{iii}$}             & 8.73 (100.00\%)             & 3.01 (100.00\%)           & 1.49 (100.00\%)           \\
\textbf{B$^{iii}$}             & 8.74 (100.18\%)             & 3.01 (99.85\%)            & 1.46 (97.70\%)            \\
\textbf{C$^{iii}$}             & 8.72 (99.93\%)              & 3.02 (100.37\%)           & 1.43 (95.75\%)            \\
\textbf{D$^{iii}$}             & 8.64 (99.03\%)              & 2.90 (96.40\%)            & 1.41 (94.45\%)            \\
\textbf{E$^{iii}$}             & 8.25 (94.49\%)              & 2.48 (82.32\%)            & 1.07 (71.77\%)            \\
\textbf{F$^{iii}$}             & 8.15 (93.35\%)              & 2.34 (77.72\%)            & 0.85 (57.19\%)            \\
\textbf{G$^{iii}$}             & 8.17 (93.62\%)              & 2.30 (76.32\%)            & 0.86 (57.40\%)           \\                        

\bottomrule
\end{tabular*}
\vspace{-0.4cm}
\begin{flushleft}
{\footnotesize{$^i$ Combination of super mirror indexes of $m=3.0$ and $m=2.5$}}\\
{\footnotesize{$^{ii}$ Combination of super mirror indexes of $m=3.0$ and $m=2.0$}}\\
{\footnotesize{$^{iii}$ Combination of super mirror indexes of $m=2.5$ and $m=2.0$}}
\end{flushleft}
\end{table}

\begin{table}[hbt!]
\caption{Flux data results of the last run of simulations. Configurations of Table \ref{M_DIF_CASES_IV} are carried out for three different ILL virtual sources, i.e., with cold, thermal, and hot neutron spectra. Cold and thermal fluxes, described respectively by the second and third columns, are given in $\times 10^9$~$n/cm^2s$, while hot results in the last column are in $\times 10^8$~$n/cm^2s$. Flux values come together with percentages between parenthesis that correspond to the ratio of each flux value next to the case A of each sequence. These percentages are just relative to cases \textbf{A} to \textbf{G} and their respective subcases, i.e., \textbf{i}, \textbf{ii}, and \textbf{iii}, and cases \textbf{A} are the standard comparison value (always with percentage $100$~$\%$). Results correspond to simulations of case \textbf{Iv} from Table \ref{tbl1}.}\label{FLUX_IV}
\begin{tabular*}{\tblwidth}{@{} LLLL@{} }
\toprule
\multirow{3}{*}{\textbf{Case}} & \textbf{Cold}               & \textbf{Thermal}          & \textbf{Hot}              \\
                               & $F_{cur}(R_{flux}^{abs})$   & $F_{cur}(R_{flux}^{abs})$ & $F_{cur}(R_{flux}^{abs})$ \\
                               & $ (\times 10^{9}   n/cm^2s)$ & $ (\times 10^{9} n/cm^2s)$ & $ (\times 10^{8} n/cm^2s)$ \\

\midrule 
\textbf{A$^{i}$}               & 9.81 (100.00\%)             & 3.54 (100.00\%)           & 1.79 (100.00\%)           \\
\textbf{B$^{i}$}               & 9.87 (100.60\%)             & 3.46 (97.64\%)            & 1.77 (99.14\%)            \\
\textbf{C$^{i}$}               & 9.84 (100.34\%)             & 3.47 (97.82\%)            & 1.76 (98.30\%)            \\
\textbf{D$^{i}$}               & 9.81 (99.99\%)              & 3.43 (96.84\%)            & 1.80 (100.74\%)           \\
\textbf{E$^{i}$}               & 9.65 (98.37\%)              & 3.18 (89.79\%)            & 1.43 (79.83\%)            \\
\textbf{F$^{i}$}               & 9.48 (96.62\%)              & 2.94 (83.10\%)            & 1.21 (67.99\%)            \\
\textbf{G$^{i}$}               & 9.47 (96.50\%)              & 2.95 (83.35\%)            & 1.21 (67.72\%)            \\
\textbf{A$^{ii}$}              & 9.85 (100.00\%)             & 3.53 (100.00\%)           & 1.79 (100.00\%)           \\
\textbf{B$^{ii}$}              & 9.87 (100.18\%)             & 3.46 (98.19\%)            & 1.79 (100.21\%)           \\
\textbf{C$^{ii}$}              & 9.68 (98.22\%)              & 3.35 (95.06\%)            & 1.81 (101.25\%)           \\
\textbf{D$^{ii}$}              & 9.57 (97.14\%)              & 3.38 (95.89\%)            & 1.79 (99.89\%)            \\
\textbf{E$^{ii}$}              & 8.92 (90.54\%)              & 2.55 (72.36\%)            & 0.95 (52.87\%)            \\
\textbf{F$^{ii}$}              & 8.64 (87.69\%)              & 2.24 (63.58\%)            & 0.66 (36.76\%)            \\
\textbf{G$^{ii}$}              & 8.59 (87.69\%)              & 2.22 (62.88\%)            & 0.67 37.59\%)             \\
\textbf{A$^{iii}$}             & 9.48 (100.00\%)             & 2.90 (100.00\%)           & 1.22 (100.00\%)           \\
\textbf{B$^{iii}$}             & 9.50 (100.27\%)             & 2.89 (99.59\%)            & 1.20 (98.73\%)            \\
\textbf{C$^{iii}$}             & 9.39 (99.06\%)              & 2.91 (100.49\%)           & 1.23 (100.84\%)           \\
\textbf{D$^{iii}$}             & 9.37 (98.90\%)              & 2.92 (100.88\%)           & 1.15 (94.65\%)            \\
\textbf{E$^{iii}$}             & 9.16 (96.63\%)              & 2.45 (84.43\%)            & 0.85 (69.55\%)            \\
\textbf{F$^{iii}$}             & 8.57 (90.39\%)              & 2.23 (77.04\%)            & 0.67 (54.84\%)            \\
\textbf{G$^{iii}$}             & 8.56 (90.31\%)              & 2.21 (76.33\%)            & 0.66 (54.45\%)           \\                        

\bottomrule
\end{tabular*}
\vspace{-0.4cm}
\begin{flushleft}
{\footnotesize{$^i$ Combination of super mirror indexes of $m=3.0$ and $m=2.5$}}\\
{\footnotesize{$^{ii}$ Combination of super mirror indexes of $m=3.0$ and $m=2.0$}}\\
{\footnotesize{$^{iii}$ Combination of super mirror indexes of $m=2.5$ and $m=2.0$}}
\end{flushleft}
\end{table}

\begin{figure}[hbt!]
\centering
\begin{subfigure}{.225\textwidth}
  \centering
  \includegraphics[width=.85\linewidth]{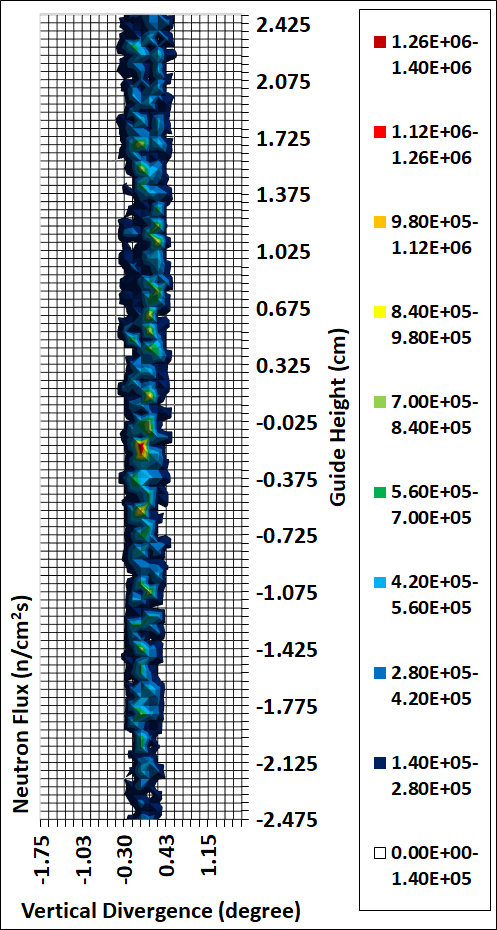}
  \caption{Case A$^{ii}$}
  \label{case_A_ii_hot}
\end{subfigure}%
\begin{subfigure}{.225\textwidth}
  \centering
  \includegraphics[width=.85\linewidth]{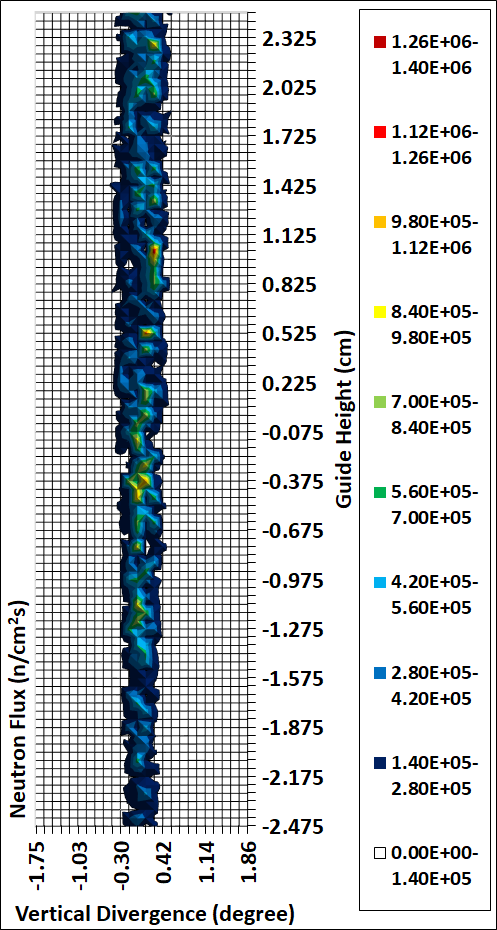}
  \caption{Case B$^{ii}$}
  \label{case_B_ii_hot}
\end{subfigure}
\begin{subfigure}{.225\textwidth}
  \centering
  \includegraphics[width=.85\linewidth]{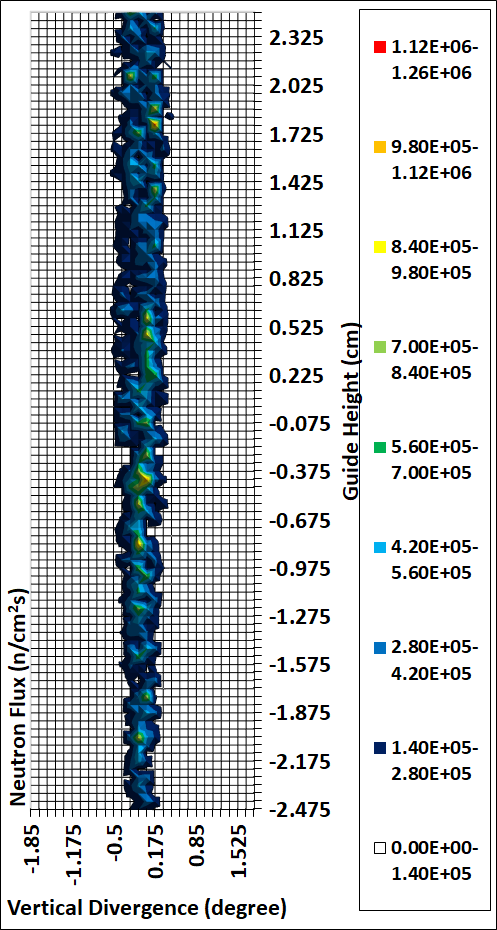}
  \caption{Case C$^{ii}$}
  \label{case_C_ii_hot}
\end{subfigure}
\begin{subfigure}{.225\textwidth}
  \centering
  \includegraphics[width=.85\linewidth]{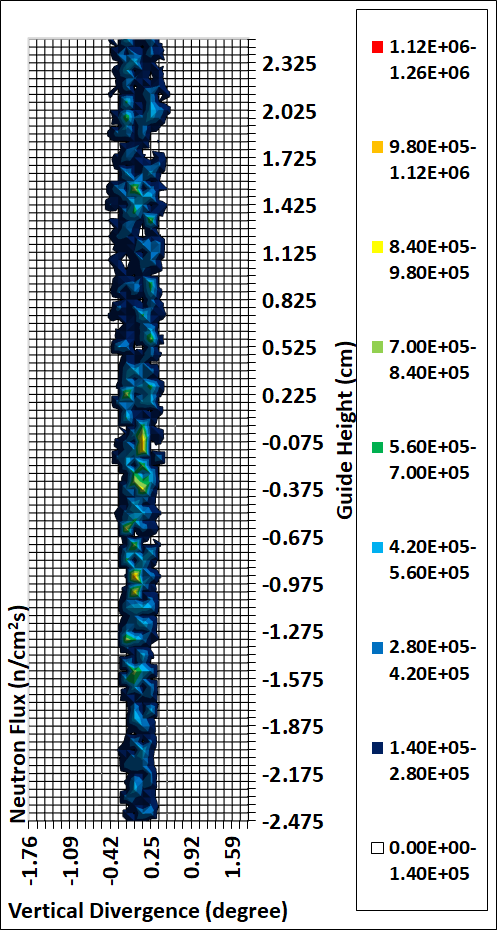}
  \caption{Case D$^{ii}$}
  \label{case_D_ii_hot}
\end{subfigure}
\begin{subfigure}{.225\textwidth}
  \centering
  \includegraphics[width=.85\linewidth]{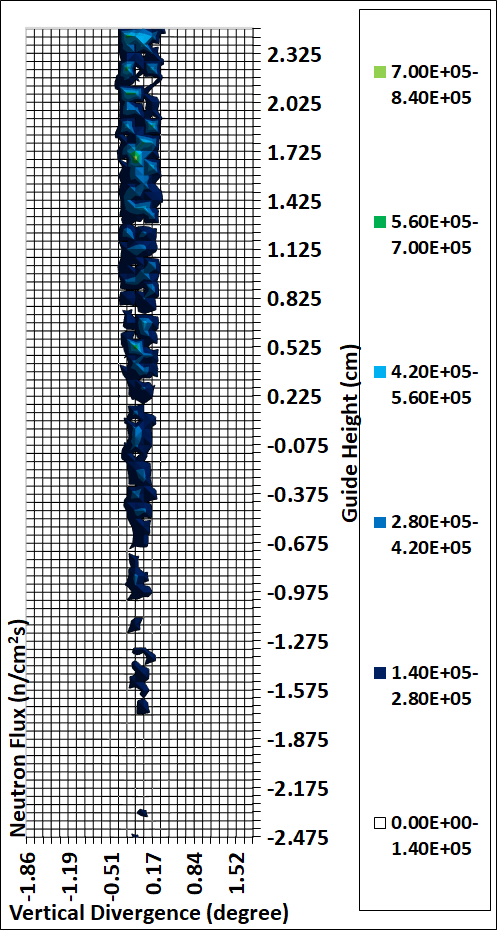}
  \caption{Case E$^{ii}$}
  \label{case_E_ii_hot}
\end{subfigure}
\begin{subfigure}{.225\textwidth}
  \centering
  \includegraphics[width=.85\linewidth]{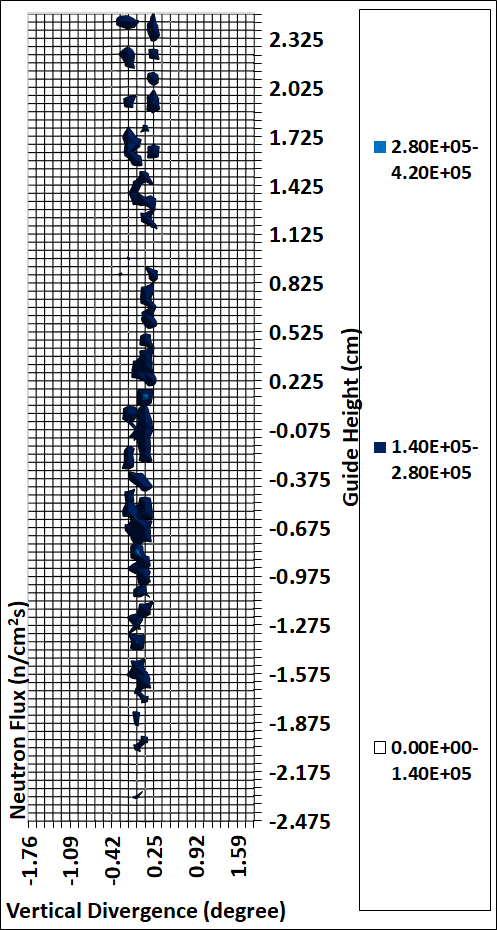}
  \caption{Case F$^{ii}$}
  \label{case_F_ii_hot}
\end{subfigure}
\caption{Acceptance diagrams at the end of the S-shaped guide system from simulations of cases $A^{ii}$, $B^{ii}$, $C^{ii}$, $D^{ii}$, $E^{ii}$, and $F^{ii}$ with ILL hot virtual source. Diagrams are obtained by considering the whole spectra, i.e., from $0$ to $20$ \AA. Ordinate and abscissa axes stand for guide width ($cm$) and vertical divergence (degrees). Minimal flux range, which is represented by white color, is fixed between $0$ and $1.40\times 10^5$~$n/cm^2s$, while other fluxes ranges come from ``cold'' colors (blue) to ``hot'' colors (red) in steps of $1.40\times 10^5$~$n/cm^2s$. }
\label{ad_diagram}
\end{figure}

To compare fluxes of different cases, i.e., \textbf{A} to \textbf{G}, percentages of each value next to case one are displayed after their values between parenthesis. Since \textbf{A} cases represent combinations of maximum coating index values, their percentages are always $100$~$\%$. Even though \textbf{A} cases are our standard, it is possible to find higher fluxes in other configurations, i.e., $>100$~$\%$, due to simulation errors. According to AD formalism and space phase tailoring, we consider configurations \textbf{D} the most promising scenario combining flux and divergence with lower coating indexes along with an S-shaped guide system. By checking Tables \ref{FLUX_II} and \ref{FLUX_IV}, it may be observed that cases \textbf{A}$^{ii}$, \textbf{B}$^{ii}$, \textbf{C}$^{ii}$, \textbf{D}$^{ii}$, \textbf{E}$^{ii}$, \textbf{F}$^{ii}$, and \textbf{G}$^{ii}$ of the ILL hot source present the most abrupt fall in neutron flux. ADs of these cases, except case G, are presented in Figure \ref{ad_diagram}. From these diagrams, we confirm that case \textbf{D} presents a similar flux to cases \textbf{A}, \textbf{B}, and \textbf{C}, but with lower coating indexes. By comparing Diagrams \ref{case_A_ii_hot} and \ref{case_D_ii_hot} it is not possible to identify phase space tailoring since the second straight guide is too short compared to the curved ones in the S-shaped guide system, i.e., each curved guide segment is about twenty-three times longer than this straight part.

The identification of the tailoring effect is easily achieved through monochromatic spectra, since every wavelength would possess a specific removal of neutrons of unwanted divergence. Nevertheless, tests with longer secondary guides have been carried out, and results from the process of phase space tailoring could be identified by checking neutron flux strip regularity of analyzed diagrams. Additional simulations are still necessary to investigate this process, but it has been left for future analysis.

\section{Conclusions}
\label{C}

In this work, three different aspects of minimal S-shaped guide systems, based on the shortest length for LoS exclusion, were investigated. The first topic consists of investigating the wavelength cutoff process according to curved guide arc ratios utilizing twelve simulation cases (\textbf{I} - \textbf{XII}) divided into three different scenarios (\textbf{a.}, \textbf{b.} and \textbf{c.}). Here, we propose a series of equations based in combinations of curved-straight and curved-curved guides. Such equations have been used on all simulated S-shaped guide systems taking into account the minimal length that guarantees LoS exclusion. Simulation results have shown that configuration \textbf{c.} is the shortest arrangement of an S-shaped guide system that excludes LoS, and it can maintain wavelength cutoff, which according to Gilles \cite{Gilles2006}, is the classic characteristic of any S-shaped guide. Thus, if wavelength cutoff is unwanted, one can still use configurations \textbf{a.} and \textbf{b.} to build guide systems. Since just configuration \textbf{c.} shows wavelength cutoff, all posterior simulations preserve such an arrangement. 
Utilizing another series of simulations of different arc ratios, configuration \textbf{c.} was explored in scenarios with different primary and secondary straight guides and consequently different arc ratios. The results allow us to verify that both curved guides should have about $65$~$\%$ of their respective characteristic wavelength to impose a cutoff on the transported neutron profile. Both runs of simulations have produced results allowing to define the last sequence of tests.

The third and last sequence of simulations explore two aspects of S-shaped guide systems. It is given that a further analysis of the scenario $c$ considering different inner coating indexes is necessary, once in the first set of simulations, all indexes were kept the same, i.e., $m=3$. From this investigation, we intend to find the relation between inner coating indexes and wavelength cutoff. Thus, the validity of another proposed equation was tested, ``normalizing'' all coating indexes at the cost of curvature value changes.  Besides, within this last set of simulations, we also intend to make a global comparison of coating index arrangements to define optimized configurations that provide maximum flux (compared with scenarios of the same index values) using lower indexes on some inner surfaces.

The former run of simulations indicates that configurations \textbf{a.} and \textbf{b.} are not able to impose wavelength cutoff on neutron profiles for twelve different S-shaped guide system arrangements. This occurs because both configurations are geometrically built based on taking the inner point of the concave surface as a limit to avoid LoS. This point possesses the same significance as any other in the trajectory that it represents according to AD formalism, however, it imposes an ideally maximum arc ratio of $50$~$\%$ (note that the connection of curved guides reduces the arc to lower values of $\Psi_C$). From this scenario, we conclude that it is impossible to have a system that minimally excludes LoS and also provides a cutoff. In this sense, we assume configuration $c$ as a standard of the minimal S-shaped guide system. Besides, using equal curvatures in \textbf{c.} allows us to possess symmetrical systems with parallel entrance and exit sections, e.g., cases \textbf{I}, \textbf{IV}, \textbf{VI}, \textbf{VII}, \textbf{X}, and \textbf{XII}.

Most simulated wavelength cutoff values of \textbf{c.} cases agree properly with theoretical values according to the present study equations, i.e., in eleven out of twelve cases, the difference between theoretical and simulated values was less than $10\%$. From these results, we have proposed another set of simulations based on a configuration with  $\rho_2=2000$~$m$ and $\rho_3=1000$~$m$. Simulations have been carried out for twenty-one different arc ratios, from $40$~$\%$ to $80$~$\%$ in  steps of $2$~$\%$. According to Equation \ref{L_P_S}, these number of ratios are the same number for different primary and secondary straight guide lengths in simulations. Results indicate a minimal arc ratio of $65$~$\%$ to an agreement of less than $5$~$\%$ between theoretical and simulated wavelength cutoff. 

The last sequence of simulations is based on exploring cases \textbf{II} and \textbf{IV} for seven distinct coating arrangements to test the wavelength cutoff modified equation and also to select an optimized scenario that guarantees the best relation between neutron transportation and proper coating surface system. All simulations have been carried out for each of the ILL sources, i.e., hot, thermal, and cold. Due to the number of simulations, we have displayed statistical information in Table \ref{STAT} and from it, it is possible to verify that there is a good agreement between theoretical and simulated wavelength cutoffs. Considering all cases and sources, we observe a concordance of $60.32$~$\%$, i.e., 76 from 126 cases and $91.27$~$\%$, i.e., 115 from 126 cases for $5$~$\%$ and $10$~$\%$ ranges between theoretical and simulated, respectively.

Here we observe that cold source results, mainly for case \textbf{II}, possess worse agreement compared to others. We believe that is due to the number of rays ($10^7$) used in present MCSTAS simulations, since wavelength cutoff occurs close to thermal and hot spectra peaks, where these profiles have more neutrons than the cold profile. Thus, there are statistically fewer neutrons to determine the wavelength cutoff for cold sources than for hot and thermal ones. Nevertheless, and considering simulation errors, we conclude that these results confirm the validity of the applied equations. Analyzing flux values, we have been able to keep configuration \textbf{D} as a standard configuration since it maintains a slightly lower difference next to configuration \textbf{A}. In this sense, it maintains the phase space tailoring and also represents an optimized configuration as to engineering and financial terms. These aspects are reiterated since the guide design geometrically offers the shortest S-shaped guide systems excluding LoS with, or without, wavelength cutoff. Last, but not least, all the formalism presented in this study allows us to project different guide systems employing curved-straight and curved-curved guides combination, depending on the design requirements.

We finally conclude that the outcome of this study may be applied in many scenarios where guide sections present misalignment or significant vertical or horizontal displacements. The former may be used in the ESS case, where installation ground gradually sinks, and total displacements are tens of centimeters. For this scenario, we presume that the wavelength cutoff and the LoS are essential aspects. The latter case of applications comprehends situations of adapting new instruments in operating facilities or transferring instruments from different center places or inter-centers. In this scenario, we intend to propose an installation of a new SANS instrument at the Brazilian IEA-R1 reactor. There, the neutron hall does not have enough space for long or large instruments, which could be solved by moving the neutron beam exit to the upper floor. The viability of this installation through S-shaped guide systems is left for future works.  


\section*{Acknowledgement}

The authors are thankful to the technical coordinator of the Brazilian Multipurpose Reactor (RMB) Project, Dr. J.A. Perrotta. A.P.S. Souza and L.P. de Oliveira also would like to thank CNPq for financial support under grant numbers 381565/2018-1 and 380183/2019-6, respectively.


\bibliographystyle{JHEP2}

\bibliography{cas-refs}

\end{document}